\documentclass[useAMS,usenatbib,usegraphicx, a4paper]{mn2e}

\def\aap{A\&A}
\def\apj{ApJ}
\def\apjs{ApJS}
\def\apjl{ApJL}
\def\mnras{MNRAS}

\def\pra{Phys.Rev.A}         
\def\prd{Phys.Rev.D}         

\newcommand{\pot}[2]{#1 \times 10^{#2}}
\newcommand{\xe}{x_{\rm e}}
\newcommand{\omegab}{{\Omega_{\rm b}h^2}}
\newcommand{\omegadm}{{\Omega_{\rm dm}h^2}}
\newcommand{\omegak}{{\Omega_{\rm K}}}
\newcommand{\logAs}{\log(10^{10} A_{\rm S})}
\newcommand{\nS}{n_{\rm S}}
\newcommand{\nT}{n_{\rm T}}
\newcommand{\nrun}{n_{\rm run}}
\newcommand{\asz}{A_{\rm SZ}}
\newcommand{\ion}[2]{{\text{{\sc #1}\,{\sc #2}}}}

\usepackage{amsmath}
\usepackage{color}

\title[Recombination uncertainties and CMB experiments]{Estimating the impact of
 recombination uncertainties on the cosmological parameter constraints from
 cosmic microwave background experiments}

\author[Rubi\~no-Mart\'{\i}n et al.]{
J.~A. Rubi\~no-Mart\'{\i}n$^{1,2}$\thanks{E-mail: 
jose.alberto.rubino@iac.es}, J. Chluba$^{3,4}$\thanks{E-mail:
 jchluba@cita.utoronto.ca}, W.~A.~Fendt$^{5}$ and B.~D.~Wandelt$^{5,6,7}$\\
$^{1}$ Instituto de Astrof\'isica de Canarias, C/V\'ia L\'actea s/n, 
   E-38200 Tenerife, Spain\\
$^{2}$ Departamento de Astrof\'isica, Universidad de La Laguna, E-38205 La Laguna,
Tenerife, Spain\\
$^{3}$ Canadian Institute for Theoretical Astrophysics, 60 St. George Street,
Toronto, ON M5S 3H8, Canada\\
$^{4}$ Max-Planck Institut f\"ur Astrophysik, Karl-Schwarzschild-Str. 1,
D-85740 Garching, Germany\\
$^{5}$ Department of Physics, UIUC, 1110 W.~Green Street, Urbana, IL
61801\\
$^6$ Department of Astronomy, UIUC, 1002 W.~Green Street, Urbana, IL
61801\\
$^7$ Center for Advanced Studies, UIUC, 912 W.~Illinois Street, Urbana, IL 61801}

\begin{document}

\date{Received **insert**; Accepted **insert**}

\pagerange{\pageref{firstpage}--\pageref{lastpage}}
\pubyear{}

\maketitle

\label{firstpage}

\begin{abstract}
We use our most recent training set for the {\sc Rico} code to estimate the
impact of recombination uncertainties on the posterior probability distributions
which will be obtained from future CMB experiments, and in particular the {\sc
  Planck} satellite.
Using a Monte Carlo Markov Chain analysis to sample the posterior distribution
of the cosmological parameters, we find that {\sc Planck} will have biases of
$-0.7$, $-0.3$ and $-0.4$ sigmas for $\nS$, $\omegab$ and $\logAs$,
respectively, in the minimal six parameter $\Lambda$CDM model, if the
description of the recombination history given by {\sc Rico} is not used. The
remaining parameters (e.g. $\tau$ or $\omegadm$) are not significantly affected.
We also show, that the cosmology dependence of the corrections to the
recombination history modeled with {\sc Rico} has a negligible impact on the
posterior distributions obtained for the case of the {\sc Planck} satellite. In
practice, this implies that the inclusion of additional corrections to existing
recombination codes can be achieved using simple cosmology-independent `fudge
functions'.

Finally, we also investigated the impact of some recent improvements in the
treatment of hydrogen recombination which are still not included in the current
version of our training set for {\sc Rico}, by assuming that the cosmology
dependence of those corrections can be neglected.
In summary, with our current understanding of the complete recombination
process, the expected biases in the cosmological parameters inferred from {\sc
  Planck} might be as large as $-2.3$, $-1.7$ and $-1$ sigmas for $\nS$,
$\omegab$ and $\logAs$ respectively, if all those corrections are not taken into
account. We note that although the list of physical processes that could be of
importance for {\sc Planck} seems to be nearly complete, still some effort has
to be put in the validation of the results obtained by the different groups.

The new {\sc Rico} training set as well as the fudge functions used for this
paper are publicly availabe in the {\sc Rico}-webpage.
\end{abstract}


%
\section{Introduction}

The cosmic microwave background (CMB) is nowadays an essential tool of
theoretical and observational cosmology. Recent advances in the observations of
the CMB angular fluctuations in temperature and polarization
\citep[e.g. ][]{WMAP5-basic,WMAP5-params} provide a detailed description of the
global properties of the Universe, and the cosmological parameters are currently
known with accuracies of the order of few percent in many cases.
The experimental prospect for the {\sc Planck} satellite \citep{Planck2006},
which was launched on May 14th 2009, is to achieve the most detailed picture of
the CMB anisotropies down to angular scales of $\ell \sim 2500$ in temperature
and $\ell \sim 1500$ in polarization. This data will achieve sub-percent
precision in many cosmological parameters.  However, those high accuracies will
rely on a highly precise description of the theoretical predictions for the
different cosmological models. Currently, it is widely recognised that the major
limiting factor in the accuracy of angular power spectrum calculations is the
uncertainty in the ionization history of the Universe \citep[see ][]{hu1995,
 Seljak2003}.

This has motivated several groups to re-examine the problem of cosmological
recombination \citep{Zeldovich1968, Peebles1968}, taking into account detailed
corrections to the physical processes occurring during hydrogen
\citep[e.g.][]{Dubrovich2005, RHS2005, Chluba2006, Kholupenko2006, Rubino2006,
  Chluba2007, Chluba2007a, Karshenboim2008, Hirata2008, Chluba2009c,
  Chluba2009b, Chluba2009, Jentschura2009, Labzowsky2009, HirataForbes2009} and
helium recombination \citep[e.g][]{Kholupenko2007, Wong2007, Switzer2008a,
  Switzer2008b, HirataSwi2008, Rubino2008, Kholupenko2008, Chluba2009d}.
Each one of the aforementioned corrections individually leads to changes in the
ionization history at the level of $\ga 0.1$\%, in such a way that the
corresponding overall uncertainty in the CMB angular power spectra exceeds the
benchmark of $\pm 3/\ell$ at large $\ell$ \citep[for more details,
  see][hereafter FCRW09]{rico}, thus biasing any parameter constraints inferred
by experiments like {\sc Planck}, which will be cosmic variance limited up to
very high multipoles.

The standard description of the recombination process is provided by the widely
used {\sc Recfast} code \citep{Seager1999}, which uses effective three-level
atoms, both for hydrogen and helium, with the inclusion of a conveniently chosen
{\it fudge factor} which artificially modifies the dynamics of the process to
reproduce the results of a multilevel recombination code \citep{Seager2000}.

The simultaneous evaluation of all the new effects discussed above make the
numerical computations very time-consuming, as they currently require the
solution of the full multilevel recombination code. Moreover, some of the key
ingredients in the accurate evaluation of the recombination history (e.g. the
problem of radiative transfer in hydrogen and the proper inclusion of two-photon
processes) are solved using computationally demanding approaches, although in
some cases semi-analytical approximations \citep[see e.g.][]{Hirata2008} might
open the possibility of a more efficient evaluation in the future.

In order to have an accurate and fast representation of the cosmological
recombination history as a function of the cosmological parameters, two possible
approaches have been considered in the literature. The first one consists of the
inclusion of additional fudge factors to mimic the new physics, as recently done
in \cite{Wong2008} (see {\sc Recfast} v1.4.2), where they include an additional
fudge factor to modify the dynamics of helium recombination.
The second approach is the so-called {\sc Rico} code (FCRW09), which provides an
accurate representation of the recombination history by using a regression
scheme based on {\sc Pico} \citep{Fendt2007a,Fendt2007}. The {\sc Rico} code
smoothly interpolates the $X_{\rm e}(z; \vec{p})$ function on a set of
pre-computed recombination histories for different cosmologies, where $z$ is the
redshift and $\vec{p}$ represents the set of cosmological parameters.

%
In this paper, we present the results for parameter estimations using {\sc Rico}
with the most recent training set presented in FCRW09.
This permits us to accurately account for the full cosmological dependence of
the corrections to the recombination history that were included in the
multi-level recombination code which was used for the training of {\sc Rico}
(see Sect.~\ref{sec:rico} for more details).
With this tool, we have evaluated the impact of the corrections on the posterior
probability distributions that are expected to be obtained for the {\sc Planck}
satellite, by performing a complete Monte Carlo Markov Chain analysis.
The study of these posteriors have shown that the impact of the cosmology
dependence is not very relevant for those processes included into the current
{\sc Rico} training set.
Therefore, by assuming that the cosmology dependence of the correction in
general can be neglected, we have also investigated the impact of recent
improvements in the treatment of hydrogen recombination (see
Sect.~\ref{sec:Lya}).
The basic conclusion is that, with our current understanding of the
recombination process, the expected biases in the cosmological parameters
inferred from {\sc Planck} might be as large as 1.5-2.5 sigmas for some
parameters as the baryon density or the primordial spectral index of scalar
fluctuations, if all these corrections to the recombination history are
neglected.
%

%
%
The paper is organized as follows. Sect.~\ref{sec:physics} describes the current
training set for {\sc Rico}, and provides an updated list of physical processes
during recombination which were not included in FCRW09.  Sect.~\ref{sec:impact}
presents the impact of the recombination uncertainties on cosmological parameter
estimation, focusing on the case of {\sc Planck}
satellite. Sect.~\ref{sec:additional} further extends this study to account for
the remaining recombination uncertainties described in Sect.~\ref{sec:physics}.
Sect.~\ref{sec:current} presents the analysis of present-day CMB experiments,
for which the effect is shown to be negligible. Finally, the discussion and
conclusions are presented in sections \ref{sec:discussion} and
\ref{sec:conclusions}, respectively.

%
\section{Updated list of physical processes during recombination}
\label{sec:physics}

\begin{figure}
\centering
\includegraphics[width=\columnwidth]{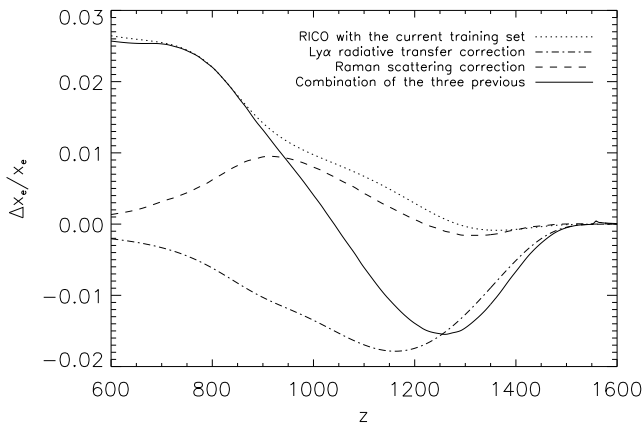}
\caption{Recombination uncertainties with respect to the standard {\sc Recfast
    v1.4.2} code. For the fiducial cosmological model, it is shown the
  correction to the recombination history which is incorporated in the new {\sc
    Rico} training set (dotted line); the correction due to Ly$\alpha$ radiative
  transfer effects \citep{Chluba2009c, Chluba2009b, Chluba2009} (dot-dashed);
  the effect of Raman scattering \citep{Hirata2008} (dashed); and the
  combination of all previous effects (solid line). }
\label{fig:remaining_xe}
\end{figure}

In this Section we provide an updated overview on the important physical
processes during cosmological recombination which have been discussed in the
literature so far.
We start with a short summary of those processes which are already included into
the current training set (FCRW09) of {\sc Rico} (Sect.~\ref{sec:rico}).
The corresponding correction to the ionization history close to the maximum of
the Thomson visibility function is shown in Fig.~\ref{fig:remaining_xe}.

We then explain the main recent advances in connection with the radiative
transfer calculations during hydrogen recombination (Sect.~\ref{sec:Lya}), which
lead to another important correction to the cosmological ionization history (see
Fig.~\ref{fig:remaining_xe}) that is not yet included into the current training
set of {\sc Rico}.
However, as we explain below (Sect.~3.4) it is possible to take these
corrections into account (Sect.~4), provided that their cosmology dependence is
negligible. Our computations show that this may be a valid approximation
(Sect.~3.4).

We end this section mentioning a few processes that have been recently addressed
but seem to be of minor importance in connection with parameter estimations for
{\sc Planck}.
Overall it seems that the list of processes that could be of importance in
connection with {\sc Planck} is nearly completed.
However, still some effort has to go in cross-validation of the results obtained
by different groups.

\subsection{The current training set for {\sc Rico}}
\label{sec:rico}
As demonstrated in FCRW09, {\sc Rico} can be used to represent the recombination
history of the Universe, accurately capturing the full cosmology dependence and
physical model of the multilevel recombination code that was used in the
computations of the {\sc Rico} training set.

For the current {\sc Rico} training set we ran our full recombination code using
a $75$-shell model for the hydrogen atom. The physical processes which are
included during hydrogen recombination are described in detail in FCRW09: the
induced 2s-1s two-photon decay \citep{Chluba2006}; the feedback of Lyman
$\alpha$ photons on the 1s-2s absorption rate \citep{Kholupenko2006}; the
non-equilibrium populations in the angular momentum sub-states \citep{Rubino2006,
 Chluba2007}; and the effect of Lyman series feedback \citep{Chluba2007a}.
For helium recombination we took into account: the spin-forbidden {He}~{\sc i}
$2^3{\rm P}_1-1^1{\rm S}_0$ transition \citep{Dubrovich2005}; and the
acceleration of helium recombination by neutral hydrogen \citep{Switzer2008a}.
Furthermore, we also updated our physical constants according to the {\sc Nist}
database\footnote{http://www.nist.gov/, 2008 May.}, including the new value of
the gravitational constant and the helium to hydrogen mass ratio
\citep{Wong2007}.

A more detailed description of the physical processes that were taken into
account in the current {\sc Rico} training set can be found in FCRW09.
This {\sc Rico} training set is now publicly available at
\verb+http://cosmos.astro.uiuc.edu/rico+.

\begin{figure}
\centering
\includegraphics[width=\columnwidth]{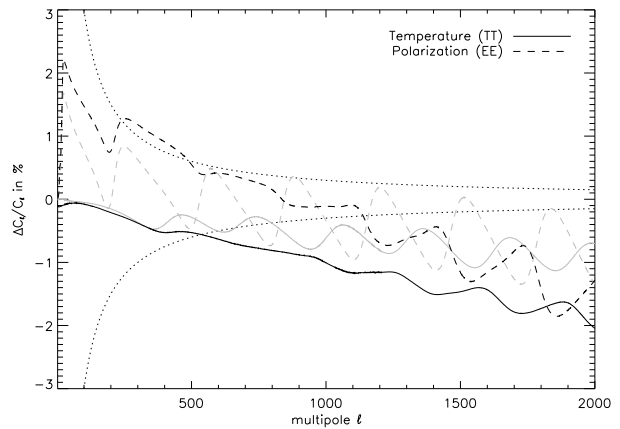}
\caption{Same as figure~\ref{fig:remaining_xe}, but for the angular power
 spectrum (APS). For the fiducial cosmological model, it is shown the
 correction to the APS due to the recombination history which is incorporated
 in the new {\sc Rico} training set (gray lines), and the correction due to the
 additional inclusion of Ly$\alpha$ radiative transfer effects
 \citep{Chluba2009c, Chluba2009b, Chluba2009} and Raman scattering
 \citep{Hirata2008} (dark lines). }
\label{fig:remaining_aps}
\end{figure}
\subsection{Updated radiative transfer calculations during hydrogen recombination}
\label{sec:Lya}
As already pointed out earlier \citep[e.g.][]{Chluba2008a} in particular a
detailed treatment of the hydrogen Lyman $\alpha$ radiative transfer problem
including two-photon corrections is expected to lead to an important additional
modification during hydrogen recombination.
An overview of the relevant physical aspects in connection with this problem was
already given in Sect. 2.3.2 and 2.3.3 of FCRW09. However, at that time the
problem was still not solved at full depth, but recently several important steps
were taken, which here we now want to discuss briefly \citep[for additional
 overview see also][]{Sunyaev2009}.

\subsubsection{Partial frequency redistribution due to line scattering}
\label{sec:Lya-recoil}
Recently, the effect of partial frequency redistribution on the Lyman $\alpha$
escape rate and the ionization history during hydrogen recombination was
independently studied in detail by \citet{Chluba2009c} and
\citet{HirataForbes2009}.
As shown by \citet{Chluba2009c}, the {\it atomic recoil} effect leads to the
dominant contribution to the associated correction in the ionization history,
and the result for this process alone seems to be in good agreement with the one
obtained earlier by \citet{Grachev2008}, leading to $\Delta N_{\rm e}/\Delta
N_{\rm e}\sim -1.2\%$ at $z\sim 900$.
Also the computations by \citet{HirataForbes2009} seem to support this
conclusion.

However, \citet{Chluba2009c} also included the effect of {\it Doppler
 broadening} and {\it Doppler boosting}\footnote{They used a Fokker-Planck
 approximation \citep[e.g.][]{Rybicki2006} for the frequency redistribution
 function.}, which was neglected in the analysis of \citet{Grachev2008}.
Doppler boosting acts in the opposite direction as atomic recoil and therefore
decelerates recombination, while the effect of Doppler broadening can lead to
both an increase in the photons escape probability or a decrease, depending on
the initial frequency of the photons \citep[see][for more detailed
  explanation]{Chluba2009c}.  The overall correction to the recombination
history due to line scattering amounts to $\Delta N_{\rm e}/\Delta N_{\rm e}\sim
-0.6\%$ at $z\sim 900$.
The results of \citet{Chluba2009c} seem to be rather similar to those of
\citet{HirataForbes2009}, however a final comparison will become necessary in
order to reach full agreement on the final correction.
In the computations presented below we will use the results of
\citet{Chluba2009c}.

\subsubsection{Two-photon transitions from higher levels}
\label{sec:2gamma}
Initially, the problem of two-photon transitions from highly excited levels in
hydrogen and helium was proposed by \citet{Dubrovich2005}. However, for hydrogen
recombination only very recently this problem has been solved convincingly by
\citet{Hirata2008} and \citet{Chluba2009b}, using two independent, conceptually
different approaches.
Also until now, \citet{Chluba2009b} took the main contribution coming from the
3d-1s and 3s-1s two-photon profile corrections into account, while
\citet{Hirata2008} also included the $n$s-1s and $n$d-1s two-photon profile
corrections for larger $n$.

In the analysis of \citet{Chluba2009b}, {\it three independent} sources for the
corrections in connection with the two-photon picture were identified (we will
discuss the other two processes in Sect.~\ref{sec:thermo} and \ref{sec:time}).
As they explain, the total modification coming from {\it purely} quantum
mechanical aspects of the problem (i.e. corrections due to deviations of the
line profiles from the normal Lorenzian shape, as pointed out by
\citet{Chluba2008a}) leads to a change in the free electron number of $\Delta
N_{\rm e}/\Delta N_{\rm e}\sim -0.4\%$ at $z\sim 1100$.
It also seems clear that the remaining small difference (at the level of $\Delta
N_{\rm e}/N_{\rm e}\sim 0.1\%-0.2\%$) between the results of \citet{Hirata2008}
and \citet{Chluba2009b} for the total correction related to the two-photon
decays from excited states is because \citet{Chluba2009b} only included the full
two-photon profiles of the 3s-1s and 3d-1s channels.
Below we will use the results of \citet{Chluba2009b} for this process.

\subsubsection{Time-dependent aspects in the emission and absorption of Lyman $\alpha$ photon}
\label{sec:time}
One of the key ingredients for the derivation of the escape probability in the
Lyman $\alpha$ resonance using the Sobolev approximation \citep{Sobolev1960} is
the {\it quasi-stationarity} of the line transfer problem. However, as shown
recently \citep{Chluba2009, Chluba2009b} at the percent-level this assumption is
not justified during the recombination of hydrogen, since (i) the ionization
degree, expansion rate of the Universe and Lyman $\alpha$ death probability
change over a characteristic time $\Delta z/z\sim 10\%$, and (ii) because a
significant contribution to the total escape probability is coming from photons
emitted in the distant wings (comparable to $10^2-10^3$ Doppler width) of the
Lyman $\alpha$ resonance.
Therefore, one has to include {\it time-dependent aspects} in the emission and
absorption process into the line transfer problem, leading to a delay of
recombination by $\Delta N_{\rm e}/N_{\rm e}\sim +1.2\%$ at $z\sim 1000$.
Below we will use the results of \citet{Chluba2009b}.

\subsubsection{Thermodynamic asymmetry in the Lyman $\alpha$ emission and absorption profile}
\label{sec:thermo}
As explained by \citet{Chluba2009b}, the largest correction related to the
two-photon formulation of the Lyman $\alpha$ transfer problem is due to the {\it
 frequency-dependent asymmetry} between the emission and absorption profile
around the Lyman $\alpha$ resonance.
This asymmetry is given by a thermodynamic correction factor, which has an
exponential dependence on the detuning from the line center, i.e. $f_\nu \propto
\exp[h(\nu-\nu_{\alpha})/kT_{\gamma}]$, where $\nu_\alpha$ is the transition
frequency for the Lyman $\alpha$ resonance.
Usually this factor can be neglected, since for most astrophysical problems the
main contribution to the number of photons is coming from within a few Doppler
width of the line center, where the thermodynamic factor indeed is very close to
unity. However, in the Lyman $\alpha$ escape problem during hydrogen
recombination contributions from the very distant damping wings are also
important \citep{Chluba2009, Chluba2009b}, so that there $f_\nu\neq 1$ has to be
included.

As explained in \citep{Chluba2008a}, the thermodynamic factor also can be
obtained in the classical picture, using the {\it detailed balance
 principle}. However, in the two-photon picture this factor has a natural
explanation in connection with the absorption of photons from the CMB blackbody
ambient radiation field \citep{Chluba2009b, Sunyaev2009}.
This process leads to a $\sim 10\%$ increase in the Lyman $\alpha$ escape
probability, and hence accelerates hydrogen recombination. For the correction to
the ionization history, \citet{Chluba2009b} obtained $\Delta N_{\rm e}/\Delta
N_{\rm e}\sim -1.9\%$ at $z\sim 1100$.
Note also that in the analysis of \citet{Chluba2009b} the thermodynamic
correction factor was included for all $n$s-1s and $n$d-1s channels with $3\leq
n \leq10$.
Below we will use the results of \citet{Chluba2009b} for this process.

\subsubsection{Raman scatterings}
\label{sec:Raman}
\citet{Hirata2008} also studied the effect of Raman scatterings on the
recombination dynamics, leading to an additional delay of hydrogen recombination
by $\Delta N_{\rm e}/N_{\rm e}\sim 0.9\%$ at $z\sim 900$.
Here in particular the correction due to 2s-1s Raman scatterings is important.
Again it is expected that a large part of this correction can be attributed to
time-dependent aspects and the correct formulation using detailed balance, and
we are currently investigating this process in detail.
In the computations presented below we will use the results of
\citet{Hirata2008} for the effect of Raman scatterings.

\subsection{Additional processes}
\label{sec:addproc}
There are a few more processes that here we only want mention very briefly
(although with this the list is not meant to be absolutely final or complete),
and which we did not account for in the computations presented here. However, it
is expected that their contribution will not be very important.

\subsubsection{Effect of electron scattering}
\label{sec:e-scatt}
The effect of {\it electron scattering} during hydrogen recombination was also
recently investigated by \citet{Chluba2009b} using a Fokker-Planck
approach. This approximation for the frequency redistribution function may not
be sufficient towards the end of hydrogen recombination, but in the overall
correction to the ionization history was very small close the maximum of the
Thomson visibility function, so that no big difference are expected when more
accurately using a scattering Kernel-approach.
Very recently \citet{Haimoud2009} showed that this statement indeed seems to be
correct.

\subsubsection{Feedback of helium photons}
\label{sec:feedbackHe}
Very recently \citet{Chluba2009d} investigated the feedback problem of helium
photons including the processes of $\gamma(\ion{He}{i})\rightarrow \ion{He}{i}$,
$\gamma(\ion{He}{i})\rightarrow \ion{H}{i}$, $\gamma(\ion{He}{ii})\rightarrow
\ion{He}{i}$ and $\gamma(\ion{He}{ii})\rightarrow \ion{H}{i}$ feedback.
They found that only $\gamma(\ion{He}{i})\rightarrow \ion{He}{i}$ feedback leads
to some small correction ($\Delta N_{\rm e}/N_{\rm e}\sim +0.17\%$ at $z\sim
2300$) in the ionization history, while all the other helium feedback induced
corrections are negligible.
This is because the $\gamma(\ion{He}{i})\rightarrow \ion{H}{i}$,
$\gamma(\ion{He}{ii})\rightarrow \ion{He}{i}$ and
$\gamma(\ion{He}{ii})\rightarrow \ion{H}{i}$ feedback processes all occur in the
{\it pre-recombinational epochs} of the considered species, where the
populations of the levels are practically in full equilibrium with the free
electrons and ions.

The $\gamma(\ion{He}{i})\rightarrow \ion{He}{i}$ feedback process was already
studied by \citet{Switzer2008a}, but the result obtained by \citet{Chluba2009d}
seems to be smaller.
However, it is clear that any discrepancy in the helium recombination history at
the 0.1\% - 0.2\% level will not be very important for the analysis of future
CMB data.

We would also like to mention, that although the $\gamma(\ion{He}{i})\rightarrow
\ion{H}{i}$, $\gamma(\ion{He}{ii})\rightarrow \ion{He}{i}$ and
$\gamma(\ion{He}{ii})\rightarrow \ion{H}{i}$ feedback processes do not affect
the ionization history, they do introduce interesting changes in the
recombinational radiation, increasing the total contribution of photons from
helium by 40\% - 70\% \citep{Chluba2009d}.

\subsubsection{Other small correction}
\label{sec:othercorrs}
Recently, several additional processes during hydrogen recombination were
discussed. These include: the overlap of Lyman series resonances caused by the
thermal motion of the atoms \citep{Haimoud2009}; quadrupole transitions in
hydrogen \citep{Grin2009}; hydrogen deuterium recombination \citep{Fung2009,
  Kholupenko2009}; and 3s-2s and 3d-2s two-photon transitions
\citep{Chluba2008a}.
All these processes seem to affect the ionization history of the Universe at a
level (well) below 0.1\%.

Furthermore, one should include the small re-absorption of photons from the
2s-1s two-photon continuum close to the Lyman $\alpha$ resonance, where our
estimates show that this leads to another $\Delta N_{\rm e}/N_{\rm e}\sim
0.1\%-0.2\%$ correction.

\subsection{Overall correction}
\label{sec:final}
Figure~\ref{fig:remaining_xe} shows the current best-estimate of the remaining
overall correction to the recombination history, while
Fig.~\ref{fig:remaining_aps} translates these corrections into changes of the
angular power spectrum.
To describe the Ly-$\alpha$ transfer effects, we adopt the curve presented in
\cite{Chluba2009c}, which includes the results of the processes investigated by
\citet{Chluba2009b} and \citet{Chluba2009}.
We also include the effect of Raman scattering on hydrogen recombination as
described in \cite{Hirata2008}.

It seems that the remaining uncertainty due to processes that were not taken
into account here can still exceed the 0.1\% level, but likely will not lead to
any significant addition anymore.
However, it is clear that a final rigorous comparison of the total result from
different independent groups will become necessary to assure that the accuracy
required for the analysis of {\sc Planck} data will be reached.
Such detailed code comparison is currently under discussion.

%
\section{Impact of recombination uncertainties on cosmological parameter estimation}
\label{sec:impact}

Here, the {\sc Rico} code together with the latest training set described above
is used to evaluate the impact of the recombination uncertainties in the
cosmological parameter constraints inferred from CMB experiments.

All our analyses use the software packages {\sc
  Camb}\footnote{http://camb.info/} \citep{camb} and {\sc
  CosmoMC}\footnote{http://cosmologist.info/cosmomc/} \citep{cosmomc}.
{\sc Camb} is used to calculate the linear-theory CMB angular power spectrum,
and has been modified here to include the recombination history as described by
{\sc Rico}.
{\sc CosmoMC} uses a Monte Carlo Markov Chain (MCMC) method to sample the
posterior distribution of the cosmological parameters from a given likelihood
function which describes the experimental constraints on the CMB angular power
spectrum.
%

The default parametrization which is included inside {\sc CosmoMC} exploits some
of the intrinsic degeneracies in the CMB angular power spectrum \citep[see
 e.g.][]{Kosowsky_params}, and uses as basic parameters the following subset:
\begin{equation}
\vec{p}_{\rm std} = \{ \omegab, \omegadm, \theta, \tau, \nS, \logAs \}. 
\end{equation}
Here, $\omegab$ and $\omegadm$ are the (physical) baryon and dark matter
densities respectively, where $h$ stands for the Hubble constant in units of
100~km~s$^{-1}$~Mpc$^{-1}$; $\theta$ is the acoustic horizon angular scale;
$\tau$ is the Thomson optical depth to reionization; and $\nS$ and $A_{\rm S}$
are the spectral index and the amplitude of the primordial (adiabatic) scalar
curvature perturbation power spectrum at a certain scale $k_0$. For the
computations in this paper, we will use $k_0=0.05$~Mpc$^{-1}$.

It is important to note that the above parametrization $\vec{p}_{\rm std}$ makes
use of an approximate formula for the sound horizon \citep{1996ApJ...471..542H},
which used in its derivation some knowledge on the recombination history
provided by {\sc Recfast}. Therefore, the use of this parameter is not
appropriate if we are changing the recombination history, as this might
introduce artificial biases. For this reason, in this paper we have modified
{\sc CosmoMC} to use a new parametrization $\vec{p}_{\rm new}$, which is defined
as
\begin{equation}
\vec{p}_{\rm new} = \{ \omegab, \omegadm, H_0, \tau, \nS, \logAs \}.
\label{eq:newpar}
\end{equation}
where we have replaced $\theta$ by $H_0$ as a basic parameter. In this way, we
still exploit some of the well-known degeneracies (e.g. that between $\tau$ and
$A_{\rm S}$), but we eliminate the possible uncertainties which may arise from
the use of $\theta$, at the expense of decreasing slightly the speed at which
the chain converges.
As we show below, both the shape of the posteriors and the confidence levels for
the rest of the parameters is practically unaffected by this modification in the
parametrization, despite of the fact that the {\sc CosmoMC} code now assumes a
flat prior on $H_0$ instead of a flat prior on $\theta$. A modified version of
the \verb!params.f90! subroutine inside {\sc CosmoMC} which uses this new
parametrization is also available in the {\sc Rico} webpage.

Finally, concerning the convergence of the chains, all computations throughout
this paper were obtained using at least five independent chains; those chains
have been run until the \citet{Gelman92} convergence criterion $R-1$ yields a
value smaller than $0.005$ for the minimal (6-parameter) case, and values
smaller than $0.02-0.2$ for those cases with a larger number of parameters.

\subsection{Impact on parameter estimates for {\sc Planck} alone}
\label{sec:planck_alone}

For the case of {\sc Planck} satellite, the mock data is prepared as follows.
We assume a Gaussian symmetric beam, and the noise is taken to be uniform across
the sky. For definiteness, we have adopted the nominal values of the beam and
pixel noise which correspond to the 143~GHz {\sc Planck} band, as described in
\cite{Planck2006}. Thus, we have $\theta_{\rm beam}= 7.1'$, $w_T^{-1} =
\sigma_{\rm noise}^2 \Omega_{\rm beam} = \pot{1.53}{-4}$~$\mu$K$^2$ and
$w_P^{-1}=\pot{5.59}{-4}$~$\mu$K$^2$, where $w_T$ and $w_P$ indicate the
intensity and polarization sensitivities, respectively.
Note that using a single frequency channel is implicitly assuming that the
remaining {\sc Planck} frequencies have been used to fully remove the foreground
contamination from this reference 143~GHz channel. For a complete discussion on
which combination of {\sc Planck} channels is more appropriate in terms of the
parameter constraints, see \cite{Colombo2009}.

The fiducial cosmological model used for these computations corresponds to the
WMAP5 cosmology \citep{WMAP5-basic}, and has parameters $\omegab = 0.02273$,
$\omegadm=0.1099$, $h=71.9$, $\tau = 0.087$, $\nS= 0.963$ and an amplitude
$\pot{2.41}{-9}$ at $k=0.002$~Mpc$^{-1}$ (or equivalently, $A_{\rm S} =
\pot{2.14}{-9}$ at $k=0.05$~Mpc$^{-1}$, or $\logAs = 3.063$).
The mock data is then produced using the cosmological recombination history as
computed with our complete multi-level recombination code for this particular
cosmology (using $n=75$ shells to model the hydrogen atom).
The shape of the likelihood function adopted here corresponds to the exact
full-sky Gaussian likelihood function given in \cite{Lewis2005}, which is
implemented in {\sc CosmoMC} using the \verb+all_l_exact+ format for the
data\footnote{See also http://cosmocoffee.info/viewtopic.php?t=231 for more
  details. }. We note that for the mock {\sc Planck} data we use not a
simulation but the actual (fiducial) angular power spectrum.

\begin{figure*}
\centering
\includegraphics[height=15cm,angle=90]{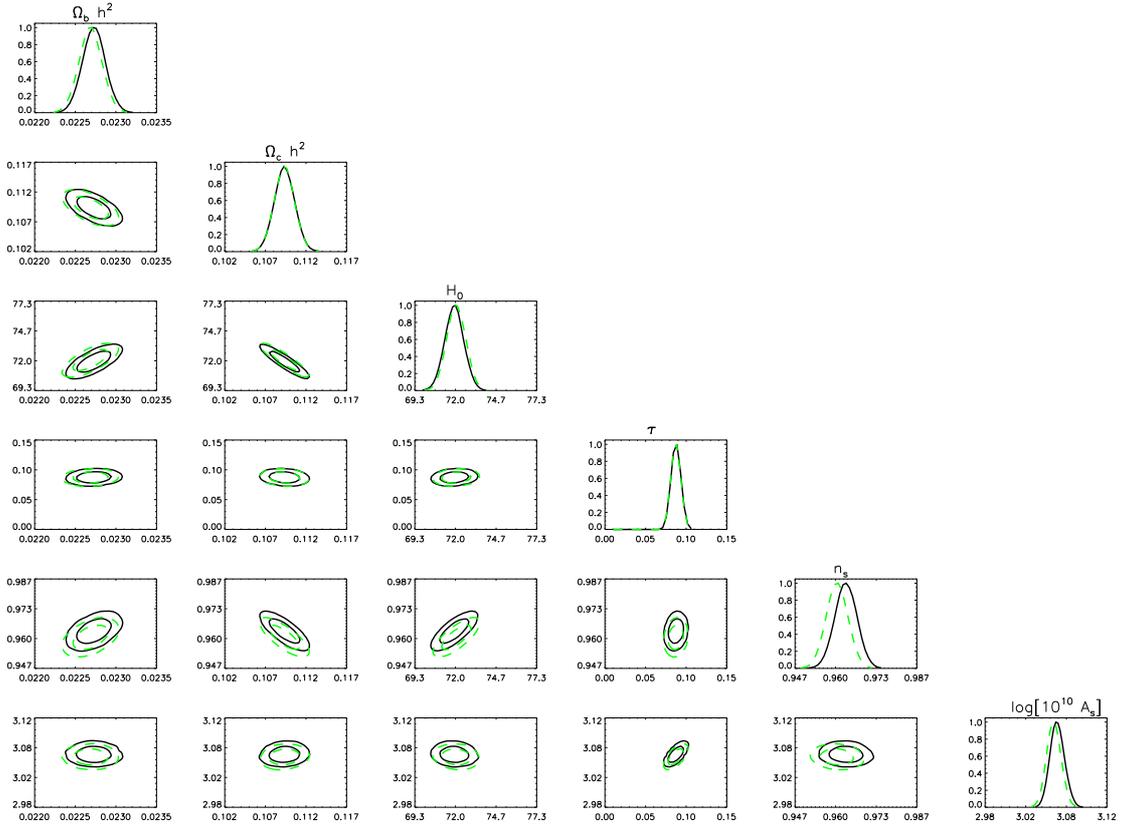}
\caption{Impact on parameter estimates for {\sc Planck} satellite, in the
  six-parameter case. Dark solid line shows the 1-D and 2-D posterior
  distributions which are obtained when the {\sc Rico} code is used to describe
  the recombination history. Light dashed solid line shows the same posteriors
  obtained with the {\sc Recfast} code. The biases are evident in $\nS$,
  $\omegab$ and $\logAs$. }
\label{fig:planck}
\end{figure*}

\begin{table*}
\centering
\caption{Impact of recombination uncertainties on the confidence limits of the
 cosmological parameters for the Planck satellite. Confidence intervals are
 derived as the 0.16, 0.5 and 0.84 points of the cumulative probability
 distribution function, in such a way that our parameter estimate is the median
 of the marginalised posterior probability distribution function, and the
 confidence interval encompasses 68 per cent of the probability. The biases
 correspond to the case of using the {\sc Rico} training set. }
\begin{tabular}{lcccc}
\hline
& With {\sc Rico} &  With {\sc Recfast} & Bias        & Fiducial \\
&                 &      (v1.4.2)       & (in sigmas) & model \\ 
\hline 
$\omegab (\times 10^2)$ & $2.273 \pm 0.014$ & $2.269 \pm 0.014$ & -0.31 &
2.273\\
$\omegadm$ & $0.1098^{+0.0013}_{-0.0012}$ & $0.1099^{+0.0012}_{-0.0013}$ & 0.06
& 0.1099\\
$H_0$ & $71.9^{+0.6}_{-0.7}$ & $72.0^{+0.7}_{-0.6}$ & 0.11 & 71.9\\
$\tau$ & $0.087 \pm 0.006$ & $0.087 \pm 0.006$ & -0.04 & 0.087\\
$\nS$ & $0.9632^{+0.0038}_{-0.0034}$ & $0.9606^{+0.0035}_{-0.0037}$ & -0.74 &
0.963\\
$\logAs$ & $3.063^{+0.009}_{-0.008}$ & $3.059 \pm 0.009$ & -0.42 & 3.063\\
\hline
\end{tabular}
\label{tab:planck}
\end{table*}

We run {\sc CosmoMC} for the case of {\sc Planck} mock data alone and the
minimal model with six free parameters. For each set of runs, we consider two
cases for the recombination history, one is the current version of {\sc Recfast}
(v1.4.2), and the second one is the {\sc Rico} code with our latest training
set.
The main result are summarized in table~\ref{tab:planck}, and the corresponding
posterior distributions are shown in Fig.~\ref{fig:planck}. We note that the
sizes of the error bars on these parameters are similar to those obtained for
{\sc Planck} by other authors \citep[e.g][]{Bond2004,Colombo2009}.
For this six parameter case, the largest biases do appear in $\nS$ ($-0.7$
sigmas), $\omegab$ ($-0.3$ sigmas) and $\logAs$ ($-0.4$ sigmas).

The sign of the correction in $\nS$, which is the parameter having the largest
bias, can be understood as follows. The physical effects to the recombination
history included in {\sc Rico} produce a slight delay of the recombination
around the peak of the visibility function, i.e. an excess of electrons with
respect to the standard computation (see Fig.~\ref{fig:remaining_xe}). This in
turn produces a slightly larger Thomson optical depth, which increases the
damping of the anisotropies at high multipoles.  In order to compensate this
excess of damping, the analysis which uses the standard {\sc Recfast} code gives
a lower value of $\nS$.

\begin{figure}
\centering
\includegraphics[width=0.9\columnwidth]{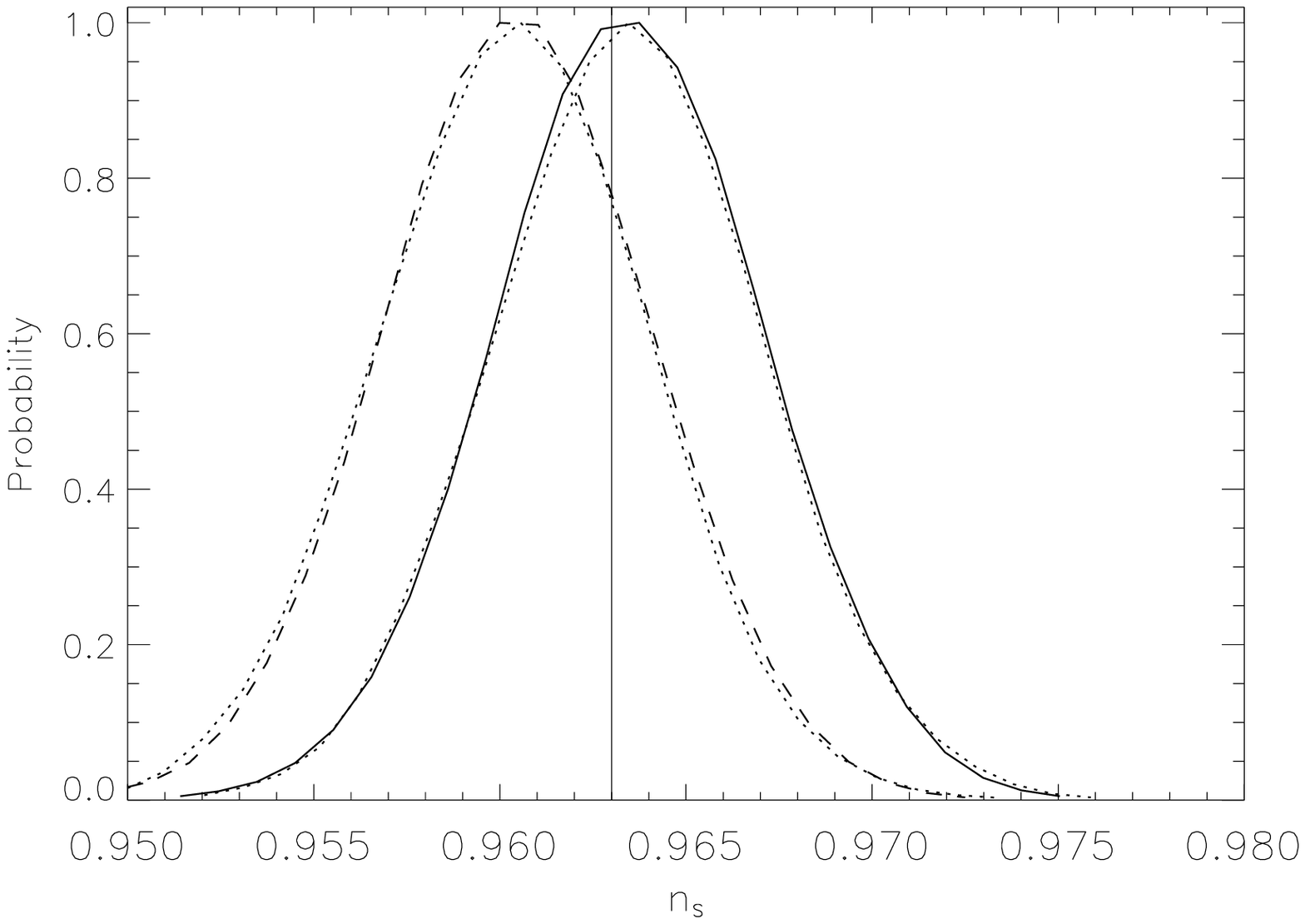}
\includegraphics[width=0.9\columnwidth]{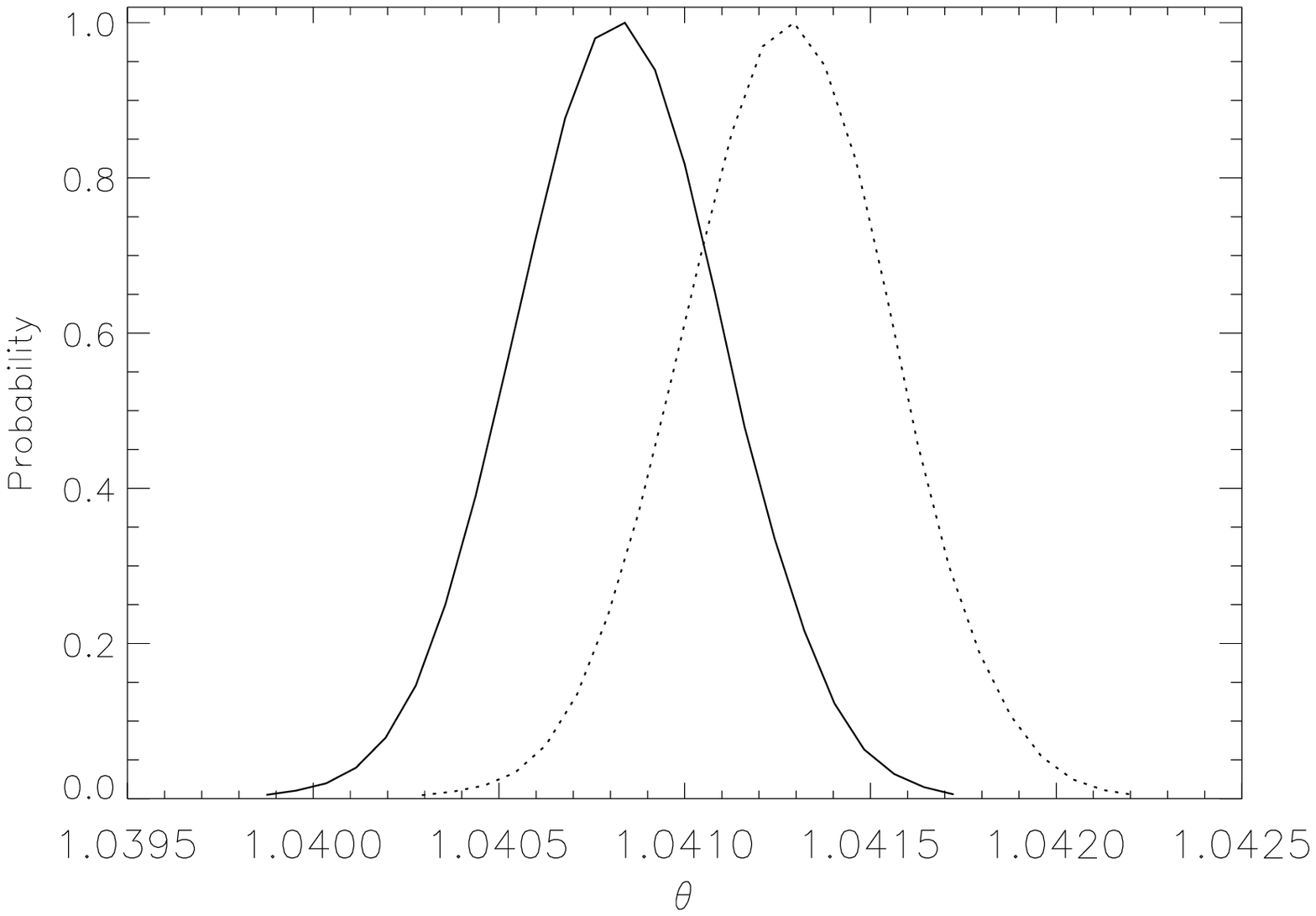}
\caption{Top: Posterior distributions for $\nS$ obtained with two different
 parametrizations. The solid line corresponds to the marginalised 1-d posterior
 distribution obtained using {\sc Rico} while the dashed line corresponds to
 the case of using {\sc Recfast}, both with the new parametrization. The dotted
 lines show in each case the posteriors obtained when the standard
 parametrization is used (i.e. $\theta$ instead of $H_0$ as fundamental
 parameter). Bottom: Bias recovered on $\theta$ parameter if the standard
 parametrization is used. This case corresponds to the same case as
 Fig.~\ref{fig:planck}, i.e. six parameter case. }
\label{fig:ns_theta}
\end{figure}

The biases on the other three parameters, i.e. $\tau$ (which is well-constrained
by large scale polarization measurements), $\omegadm$ and $H_0$, are negligible
(i.e. less than 0.1 sigmas). However, if the analysis is repeated with the
standard parametrization using $\theta$ as basic parameter, then we would find a
$+1.8$ sigma bias in $\theta$, while for the rest of the parameters the
posteriors remain the same as before. For illustration of these facts,
Fig.~\ref{fig:ns_theta} shows a comparison of the posteriors for $\nS$ obtained
with the two different parametrizations, as well as the posteriors obtained for
$\theta$ if using the standard parametrization.

Finally, we have checked that the inclusion of lensed angular power spectra on
the complete procedure modifies neither the shape of the posteriors nor the
biases. Therefore, for the rest of the paper we perform the computations in the
case without lensing. This decreases the computational time by a significant
factor.

\subsection{Importance of the cosmology dependence of the correction}
\label{sec:corr_fac}
\begin{figure}
\centering
\includegraphics[width=0.48\columnwidth]{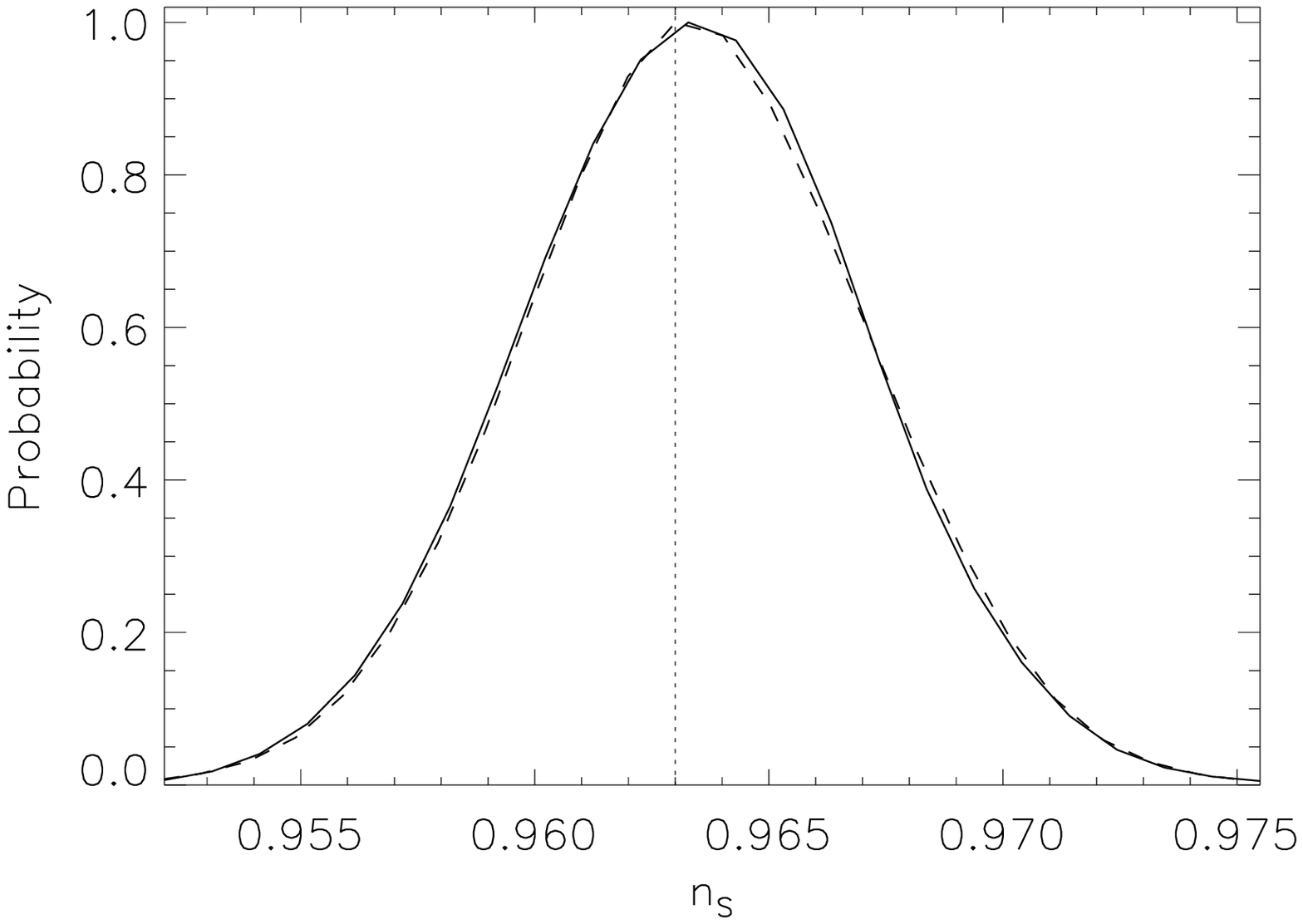}%
\includegraphics[width=0.48\columnwidth]{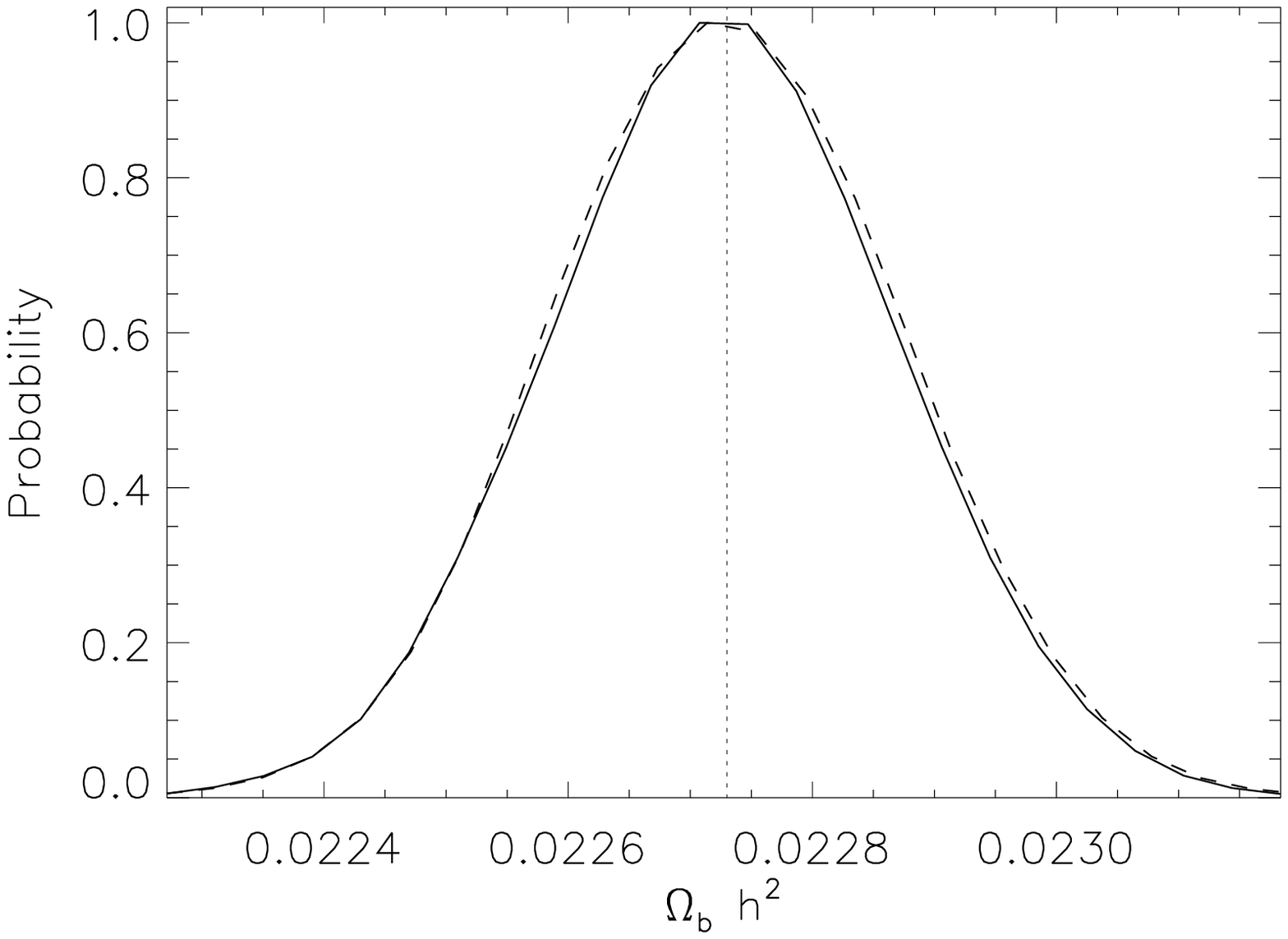}
\includegraphics[width=0.48\columnwidth]{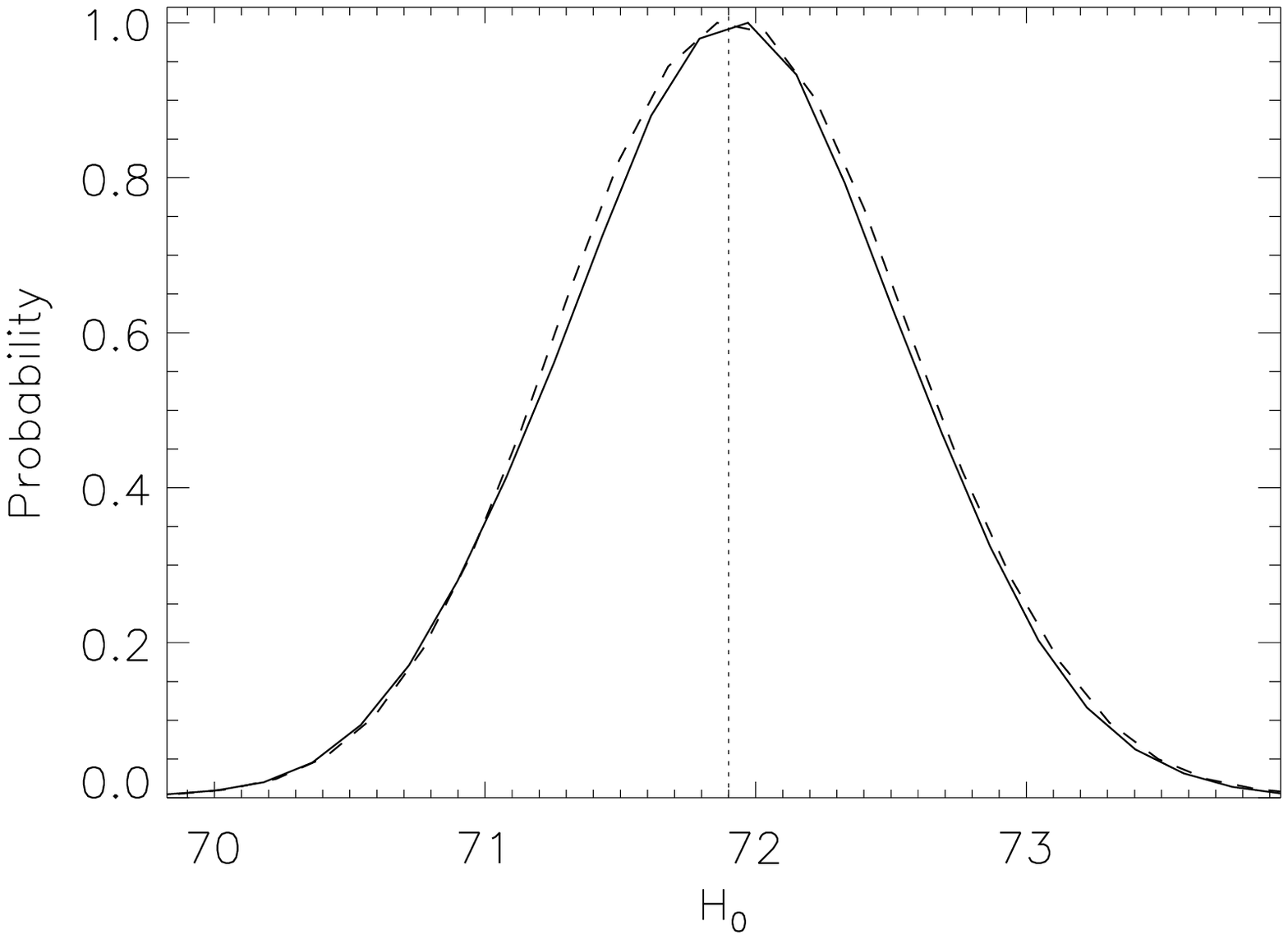}%
\includegraphics[width=0.48\columnwidth]{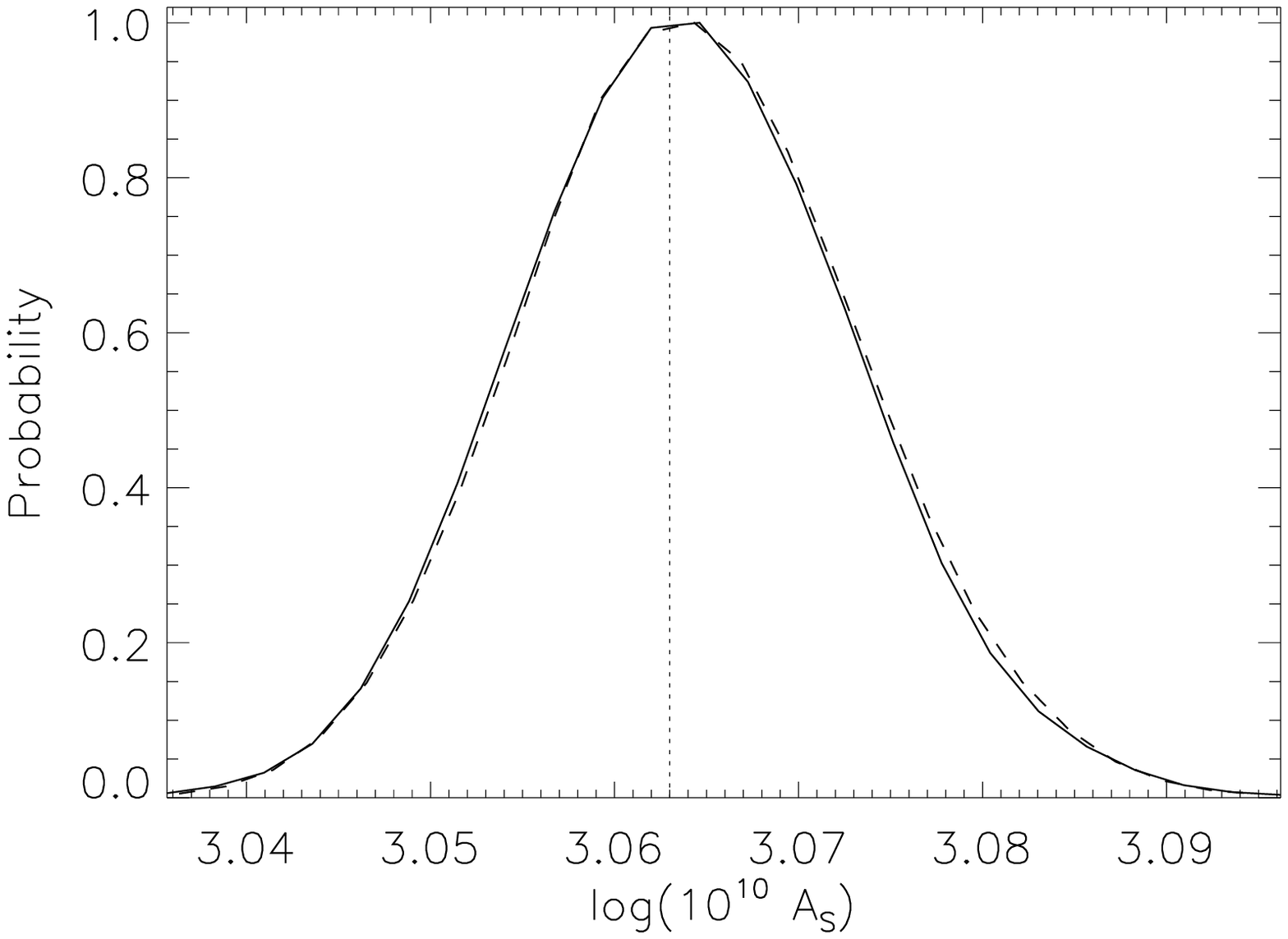}
\caption{Comparison of the 1-D posterior distributions obtained with the full
  {\sc Rico} code and the new training set (solid lines), with those obtained
  with the simplified description given in Eq.~\eqref{eq:approx} (dashed
  lines). For this case, the fiducial model which has been used to compute
  $f(z)$ is taken to be the same as the fiducial cosmology. In all four panels,
  the vertical dotted line shows the value of the fiducial model.}
\label{fig:approx}
\end{figure}

One of the questions that can be explored with {\sc Rico} and the new
training-set is the importance of the cosmology dependence of the corrections to
the recombination history. In order to obtain a simplified description of the
recombination history, it is important to evaluate if the cosmology dependence
of the corrections plays a significant role in determining the final shape of
the posteriors.
To explore this issue, we have modified the standard {\sc Recfast} code by
introducing the following (cosmology independent) correction:
\begin{equation}
\xe^{\rm new}(z ; \{ {\rm cosmology}\}) = \xe^{\rm RECFAST}(z ; \{{\rm cosmology}\}) f(z)
\label{eq:approx}
\end{equation}
where this function $f(z)$ is computed as 
\begin{equation}
f(z) = 1 + \frac{\Delta \xe}{\xe}
\label{eq:f_z}
\end{equation}
for a certain fiducial cosmological model.
By introducing this modification inside {\sc CosmoMC}, we have compared the
posterior distributions obtained in this case with those obtained using the full
{\sc Rico} code with the new training set. The results of this comparison are
shown in Figure~\ref{fig:approx}. The fact that the differences in the posterior
are negligible indicates that there is no significant cosmological dependence in
the corrections to the ionization history included in the {\sc Rico} training
set.
Therefore, and for the case of the {\sc Planck} satellite, one can in principle
use cosmology independent `fudge functions', as the one presented in
Eq.~\eqref{eq:f_z}, to accommodate additional corrections to the recombination
history.

As a final check in this section, we have explored the sensitivity of the fudge
function $f(z)$ to the cosmological model which is used as fiducial model for
the full computation of the recombination history.
Taking as a reasonable range of variation the two-sigma confidence interval
which is obtained of the analysis of WMAP5 data \citep{Dunkley2009}, we have
compared the $f(z)$ functions obtained from cosmological models which differ
two-sigmas with respect to the actual fiducial model which is used in this
paper. The result is that the changes in $f(z)$ with respect to the solid curve
presented in Fig.~\ref{fig:remaining_xe} are below one percent, thus giving a
negligible correction to the main effect included in $f(z)$.

%
\section{Estimating the corrections from the remaining recombination uncertainties}
\label{sec:additional}

Although a full recombination code which includes all the physical effects
discussed in Sect.~\ref{sec:physics} is still not available, there is a good
agreement in the community about the list of relevant physical processes that
have to be included. Moreover, all those effects have been already discussed in
the literature by at least one group, so we have estimates of the final impact
of these corrections on the recombination history (see
Fig.~\ref{fig:remaining_xe}).

\begin{figure*}
\centering
\includegraphics[width=1.3\columnwidth,angle=90]{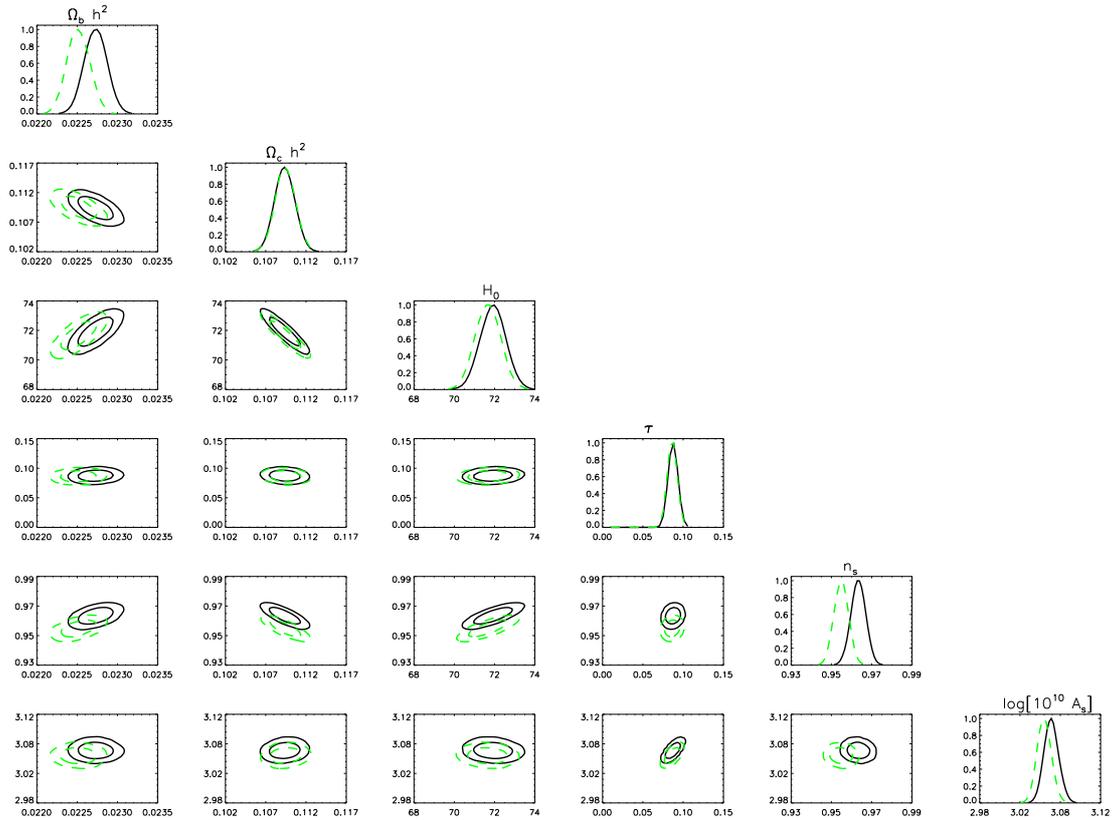}
\caption{Posterior distributions obtained with the remaining recombination
  uncertainties (see text for details). Solid line corresponds to the posteriors
  using the correct description of the recombination history, while the dashed
  lines represent the case in which the current description ({\sc Recfast
    v1.4.2}) is used. }
\label{fig:remaining_params}
\end{figure*}

Based on these estimates, we have quantified the impact of future corrections to
the recombination history using the approximation described in
Eq.~\eqref{eq:approx}, where $f(z)$ is taken from Fig.~\ref{fig:remaining_xe},
as described in Eq.~\eqref{eq:f_z}.
Using this function, we have obtained the posterior distributions for the same
case discussed above (nominal {\sc Planck} satellite sensitivities and a six
parameter analysis). The basic results are shown in Table~\ref{tab:planck2} and
Fig.~\ref{fig:remaining_params}.
The biases in the different parameters increase very significantly, as one would
expect from the inspection of Fig.~\ref{fig:remaining_xe}, specially for $\nS$
($-2.3$ sigmas), $\omegab$ ($-1.65$ sigmas) and $\logAs$ ($-1$ sigmas).
Therefore, as pointed out by several authors \citep[see e.g.][]{Chluba2008a,
  Hirata2008}, the detailed treatment of the hydrogen Lyman $\alpha$ radiative
transfer problem constitutes the most significant correction to our present
understanding of the recombinational problem. If this effect is not taken into
account when analysing {\sc Planck} data, the final constraints could be
significantly biased.

\begin{table*}
\centering
\caption{Estimate of the impact of remaining recombination uncertainties on the
 confidence limits of the cosmological parameters for the {\sc Planck}
 satellite. Confidence intervals are derived as in table~\ref{tab:planck} so
 the confidence interval encompasses 68 per cent of the probability.}
\begin{tabular}{lcccc}
\hline
& With $f(z)$ &  With {\sc Recfast} & Bias        & Fiducial \\
& from Fig.~\ref{fig:remaining_xe}  &   (v1.4.2)     & (in sigmas) & model \\ 
\hline 
$\omegab (\times 10^2)$ & $2.274^{+0.014}_{-0.015}$ & $2.250^{+0.015}_{-0.013}$
& -1.65 & 2.273\\
$\omegadm$ & $0.1098^{+ 0.0013}_{- 0.0012}$ & $0.1098^{+0.0013}_{-0.0012}$ &
0.02 & 0.1099\\
$H_0$ & $71.9 \pm 0.6 $ & $71.7 \pm 0.6$ & -0.41 & 71.9\\
$\tau$ & $0.087 \pm 0.006$ & $0.086 \pm 0.006$ & -0.18 & 0.087\\
$\nS$ & $0.963^{+0.0037}_{-0.0036}$ & $0.955^{+ 0.0035}_{- 0.0038}$ & -2.27 &
0.963\\
$\logAs$ & $3.064 \pm 0.009$ & $3.055 \pm 0.009$ & -0.99 & 3.063\\
\hline
\end{tabular}
\label{tab:planck2}
\end{table*}

\subsection{Extended cosmological models}
\label{sec:extendedmodels}

In this subsection we describe to what extend the full set of corrections to the
recombination history may affect the cosmological constraints on some extensions
to the (minimal) six-parameter model which was used in this work. Throughout
this subsection, we compare the complete recombination history (which includes
the additional corrections shown in Fig.~\ref{fig:remaining_xe}) with the
constraints that would be inferred using {\sc Recfast v1.4.2}.

\subsubsection{Tensor perturbations}

We first consider the case of including tensor perturbations in addition to the
previous model. For these computations, we consider an 8-parameter model, by
including the spectral index of the primordial tensor perturbation ($\nT$) and
the tensor-to-scalar ratio $r$, in addition to the parameters in
equation~\ref{eq:newpar}. As pivot scale, we are using $k_0=0.05$~Mpc$^{-1}$.

\begin{figure}
\centering \includegraphics[width=0.9\columnwidth]{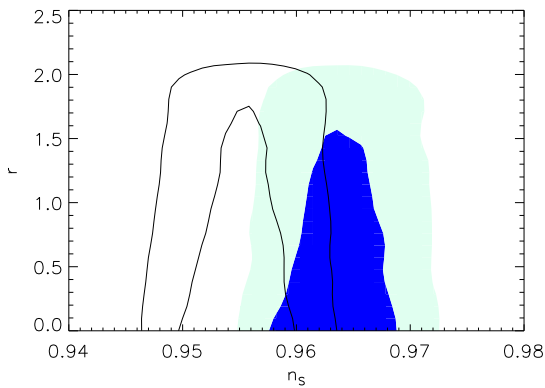}
\caption{Biases on the two dimensional marginalised constraints (68\% and 95\%)
 on inflationary parameters $r$-$\nS$. Shaded contours represent the
 constraints inferred with the complete recombination history, while the solid
 lines show the constraints using {\sc Recfast v1.4.2}. See text for details. }
\label{fig:tensors}
\end{figure}

Fig.~\ref{fig:tensors} shows the impact of recombination uncertainties on the
$r$-$\nS$ plane. The constraints on $r$ are determined by large-scale
information, and therefore the modifications of the recombination history do not
bias this parameter. However, due to the important bias on $\nS$, the 2-D
contours on this plane are significantly shifted.

\subsubsection{Scale dependence of spectral index}

\begin{figure}
\centering
\includegraphics[width=0.9\columnwidth]{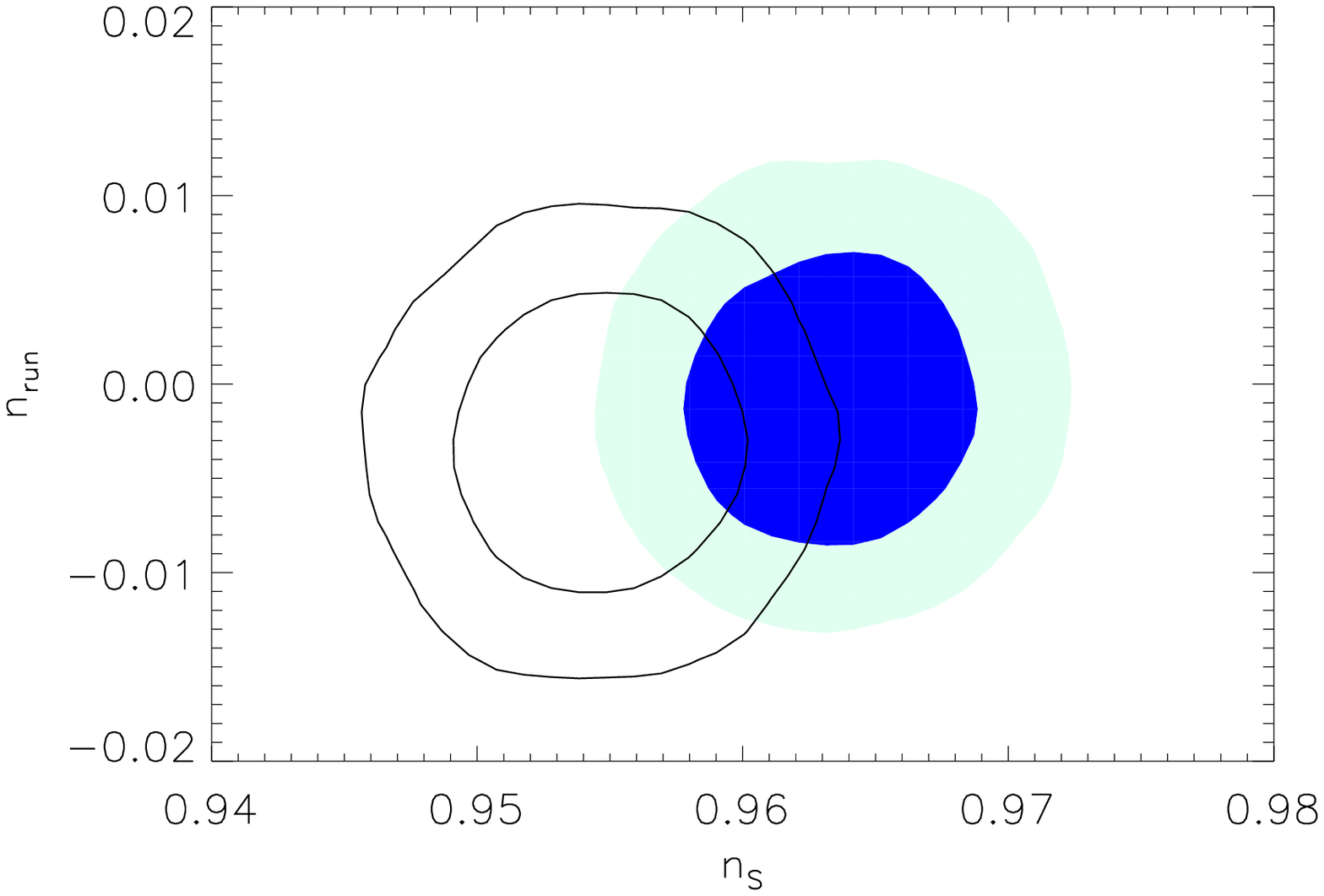}
\caption{Same as Fig.~\ref{fig:tensors}, but in the $\nS$-$\nrun$ plane. }
\label{fig:running}
\end{figure}

The running of the spectral index is a possible extension to the simple
$\Lambda$CDM model which is under debate in the literature in light of WMAP
observations \citep[see e.g.][which gives $\nrun = d \nS / d\ln k = -0.055 \pm
 0.030$]{Spergel2007}.
Fig.~\ref{fig:running} illustrate the impact of the recombination uncertainties
on the determination of the running of the spectral index. For this computation,
we considered a 7-parameter model by adding $\nrun = d \nS / d\ln k$, which
again is computed at the same pivot scale of $k_0=0.05$~Mpc$^{-1}$.
When studying the one-dimensional posterior distributions for all parameters, we
find that the inclusion of $\nrun$ does not affect the shape of the rest of the
posteriors. The bias on $\nrun$ due to recombination uncertainties is not very
significant (changes from $\nrun = -0.0012 \pm 0.0050$ to $\nrun = -0.0034 \pm
0.0050$, i.e. $0.4$ sigmas) but has to be taken into account.

For completeness, we have run a 9-parameter case, in which we allow to vary
simultaneously $r$, $\nT$ and $\nrun$ in addition to the other six parameters.
In this case, we have checked that the contours shown in Fig.~\ref{fig:tensors}
and Fig.~\ref{fig:running} are not affected by the inclusion of the other
parameter.

\subsubsection{Curvature}

\begin{figure*}
\centering
\includegraphics[width=1.3\columnwidth,angle=90]{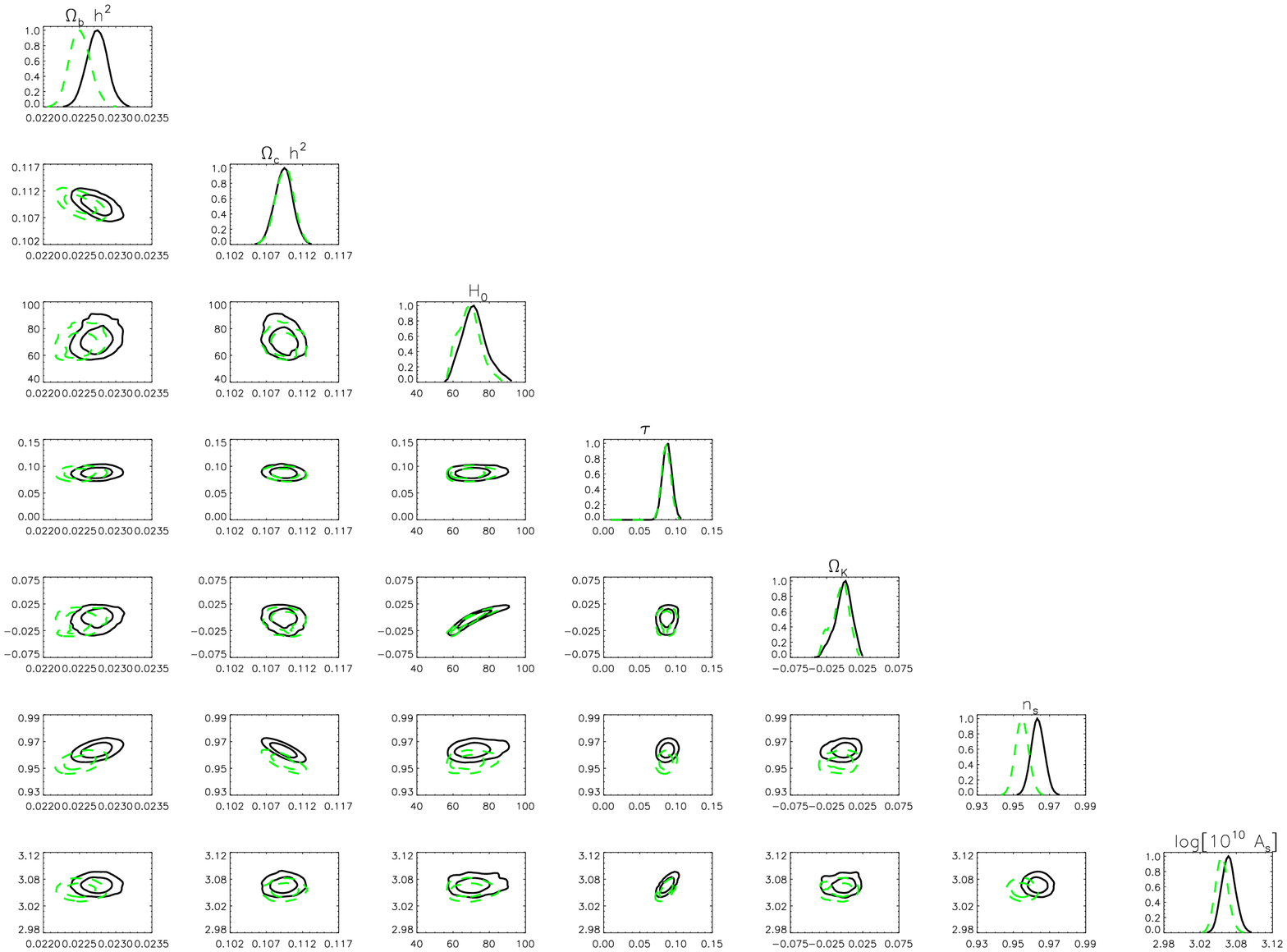}
\caption{Posterior distributions obtained with the remaining recombination
  uncertainties (see text for details) for the case of seven parameters, adding
  the curvature $\omegak$. Solid line corresponds to the posteriors using the
  correct description of the recombination history, while the dashed lines
  represent the case in which the current description ({\sc Recfast v1.4.2}) is
  used. }
\label{fig:omk}
\end{figure*}

Another possible extension of the minimal six-parameter model is to constrain
simultaneously the spatial curvature, $\omegak$. The inclusion of this
additional parameter introduces a practical complication, since in our new
parametrization $\vec{p}_{\rm new}$ (Eq.~\ref{eq:newpar}), the $H_0$ parameter
is highly correlated with $\omegak$.
Fig.~\ref{fig:omk} presents the posterior distributions for this case, in which
the degeneration between $H_0$ and $\omegak$ is clearly visible.  There are two
things to note.
First, there is no significant bias to $\omegak$ due to the inclusion of
recombination uncertainties, as one would expect since this parameter is mainly
constrained by information at angular scales around the first Doppler peak.
Second, the shape of the remaining posteriors and the biases to the parameters
are not significantly affected by the inclusion of this additional parameter.

\subsubsection{Residual SZ clusters/point source contributions}

One common extension of the minimal model is the inclusion of some parameters
describing the residual contribution of Sunyaev-Zeldovich (SZ) clusters or point
sources which are left in the maps after the component separation processes. For
the case of {\sc Planck} satellite, these are known to be the major contaminants
at small angular scales \citep[see e.g.][]{Leach2008}.

For illustration, here we will consider the case of the residual SZ cluster
contribution. One of the simplest parametrizations of the SZ contribution to the
angular power spectrum is to used a fixed template from numerical simulations,
and fit for the relative amplitude by using an additional parameter,
$\asz$. This approach is similar to the one used in \cite{Spergel2007}, where
they parametrize the SZ contribution by the model of \cite{KS2002} but allowing
for a different normalization through the $\asz$ parameter.

%
%

When the parameter constraints are infered with the inclusion of this additional
parameter, we find that (neglecting the intrinsic bias due to the degeneracy
between $\nS$ and $\asz$) the relative bias is practically not affected. In
particular, we obtain a bias of -1.42, -0.44, -2.09 and -1.03 sigmas for
$\omegab$, $H_0$, $\nS$ and $\logAs$, respectively. Those numbers are comparable
to the net biases which have been obtained for the minimal model in
Table~\ref{tab:planck2}.

\section{Impact on parameter estimates using present-day CMB experiments}
\label{sec:current}

One would expect that the order of magnitude of the corrections to the
recombination history discussed in previous sections (at the level of 1\%-2\%)
would have a negligible impact on the parameters constraints that we would infer
from present-day CMB experiments. As shown in FCRW09, the changes introduced in
the power spectra (both temperature and polarization) are significant at high
multipoles, in the sense that they are larger than the benchmark level estimated
as $\pm 3/\ell$ \citep[see][]{Seljak2003}.

\begin{figure*}
\centering 
\includegraphics[height=15cm,angle=90]{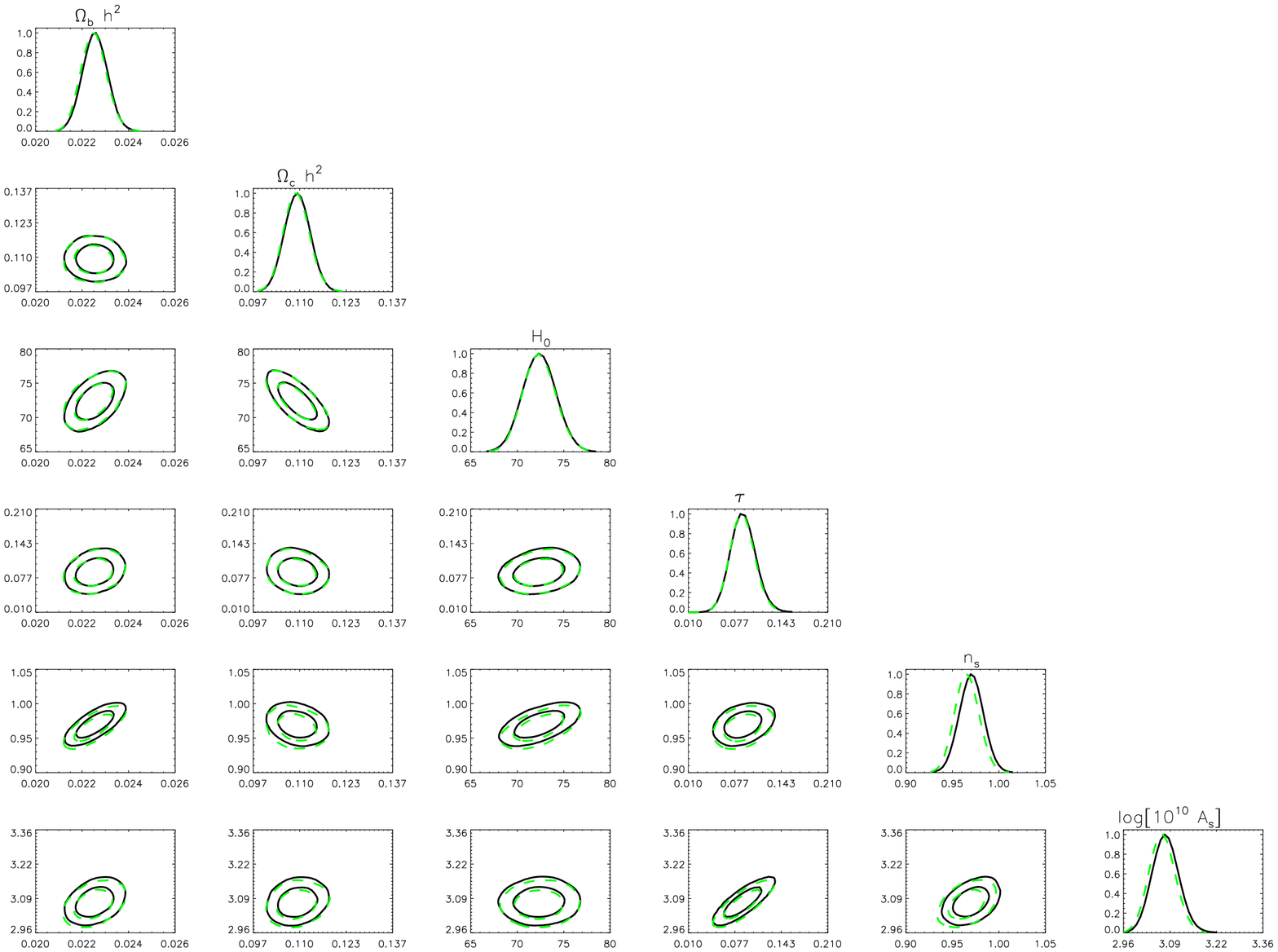}
\caption{Impact on parameter estimates for present day data
  (WMAP5+ACBAR+CBI+B03), together with SDSS LRGs. For this minimal six-parameter
  case, there are no obvious biases on the parameters, although there is a small
  difference in the shape of the posterior for $\nS$. As in
  Fig.~\ref{fig:remaining_params}, solid lines represent the posterior
  distributions obtained when our most complete description of the recombination
  process is used, while dashed lines use the current description ({\sc Recfast}
  v1.4.2). }
\label{fig:wmap}
\end{figure*}

To quantify this fact, we have obtained the posterior distributions for the case
of the minimal model with six free parameters, combining the CMB information
from WMAP5 \citep{WMAP5-basic}, ACBAR \citep{ACBAR2007}, CBI \citep{cbipol} and
Boomerang \citep{B03-TT,B03-EE}, together with measurements on the linear matter
power spectrum based on luminous red galaxies from SDSS-DR4 \citep{SDSS-LRG}.
Figure~\ref{fig:wmap} presents the results for the case of using the standard
{\sc Recfast} recombination history, together with the case of using our most
complete description of the recombination history, as presented in the previous
section. As expected, the modifications on the shape of the posteriors are very
small and no biases are seen in the parameters except for $\nS$ and $\logAs$,
which are slightly biased.
Our analysis including the full description of the recombination history gives
$\nS = 0.970 \pm 0.013$ and $\logAs = 3.075 \pm 0.038$, while the result using
{\sc Recfast v1.4.2} gives $\nS = 0.967^{+0.013}_{-0.012}$ and $\logAs =
3.066^{+0.038}_{-0.036}$. In other words, this is a $\sim -0.25$ and $\sim
-0.22$ sigma bias on $\nS$ and $\logAs$, respectively. For completeness, we have
run also the MCMC for the case of using WMAP5 data alone. In that case, the bias
decreases to $\la -0.15$ sigmas for those two parameters.

%
\section{Discussion}
\label{sec:discussion}

In this section, we now discuss the results presented in this paper focusing in
three particular aspects.
On one hand, we discuss the robustness of our results against possible
modifications of the physical description of the recombination process.
Second, we also consider the dependence of the obtained biases if additional
parameters are included in the MCMC analysis.
Finally, we discuss the possible impact of recombination uncertainties on the
results obtained from other cosmological probes different from CMB anisotropies.

\subsection{Dependence of the results on the description of the recombination process}

As discussed above (Sect.~\ref{sec:physics}), there is a wide agreement in the
community about the list of physical processes which should be included in the
description of the cosmological recombination process. In many cases, these
physical processes have been treated separately by at least two separate groups,
and the agreement on the signs and amplitudes of the corrections is excellent in
most of the cases \citep[e.g. see the compilation of uncertainties in the
  physics of recombination in Table~2.1 of][]{WongThesis}. Although an agreement
at the level of $\la 0.1$\% is still not reached, we are almost there, as the
remaining uncertainties seem to be at the level of 0.1\%-0.3\% between the
different groups.
In this sense, one would expect that a code which includes self-consistently all
those processes should obtain essentially the same biases that have been
described in Sect.~\ref{sec:additional}, and have been reported in
Table~\ref{tab:planck2}.

However, apart from the processes described in section~\ref{sec:addproc}, there
are still some possible uncertainties which might lead to measurable biases on
the cosmological parameters.
Below we now briefly address them.

\subsubsection{Hydrogen recombination}
\begin{figure}
\centering
\includegraphics[width=0.9\columnwidth]{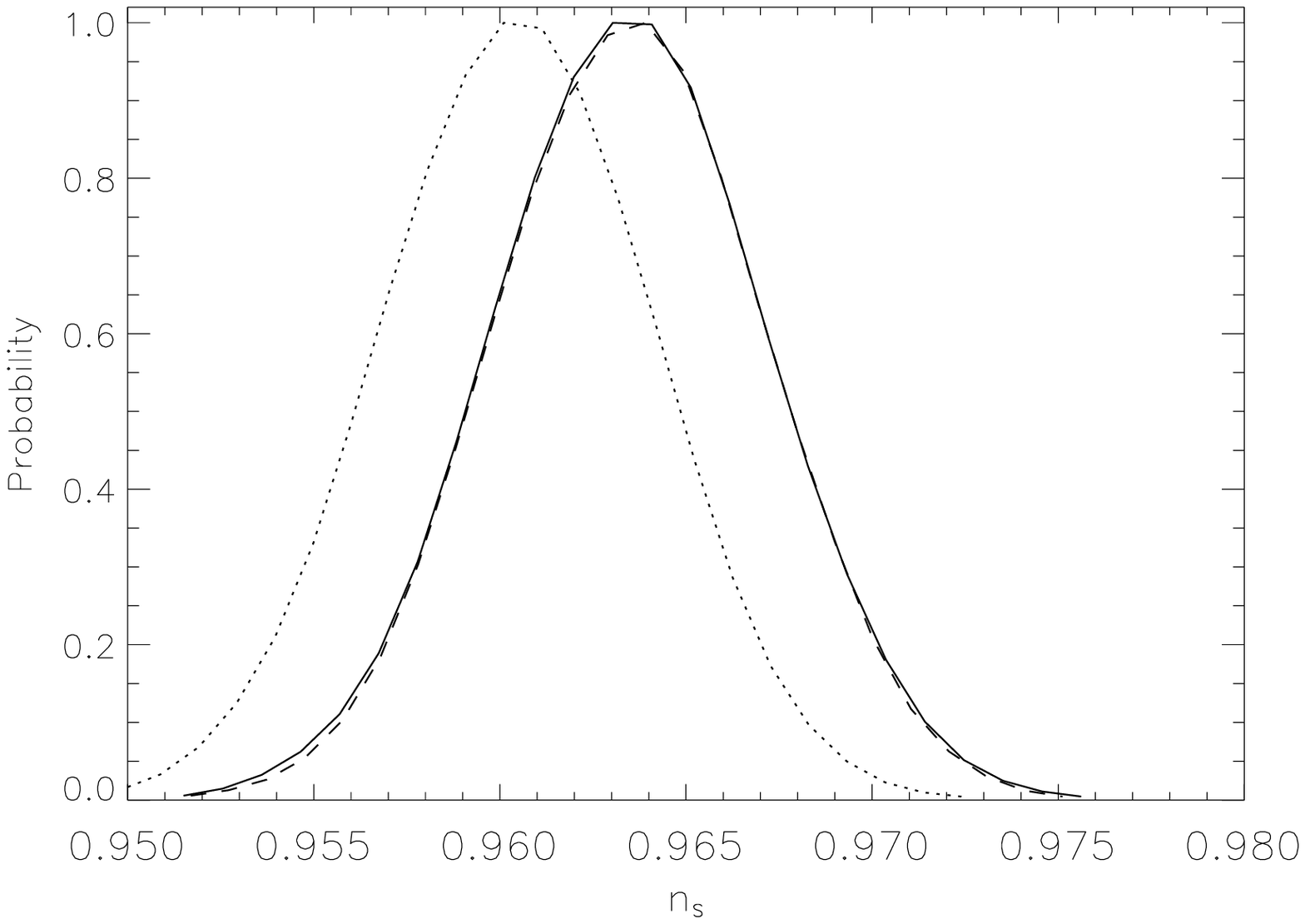}
\caption{Bias on the $\nS$ parameter due to the inclusion of additional number
  of shells in the description of the hydrogen atom. The black solid line has
  been obtained from a fiducial model with $n=110$ hydrogen shells. The dashed
  line corresponds to the posterior distribution recovered using the {\sc Rico}
  code only, with a training set based on $n=75$ shells. The dotted line
  corresponds to the posterior distribution recovered using the current {\sc
    Recfast} code. Note that we did not include for this computation the
  corrections due to Ly-$\alpha$ radiative transfer and Raman scattering. }
\label{fig:ns_fullcode}
\end{figure}

%
One effect which might lead to additional biases on the cosmological parameters
is the inclusion of very high-$n$ states in the cosmological hydrogen
recombination. The computations in this paper are based on a training set which
uses $n=75$ shells to describe the hydrogen atom. To explore the dependence of
higher number of shells, we have repeated the standard six-parameter computation
for the mock {\sc Planck} data presented in Sect.~\ref{sec:planck_alone} and
Sect.~\ref{sec:additional}, but taking as a reference model the one computed
using $n=110$ hydrogen shells, and trying to recover it with {\sc Rico} (which
uses $n=75$ hydrogen shells).
The recovered posteriors using the {\sc Rico} code in this case do not show {\it
  any appreciable} bias in any of six parameters of the minimal model. This is
illustrated in Fig.~\ref{fig:ns_fullcode}, where we present the posterior
distribution for the $\nS$ parameter, which is the one having the largest bias.

\subsubsection{Helium recombination}

%

As mentioned in Sect.~\ref{sec:addproc}, and discussed in FCRW09, we do not
expect major changes in our current understanding of the helium recombination,
at a level which might be relevant for the computation of CMB anisotropies.
This statement assumes that the uncertainties in the modelling of the helium
atom are well controlled, but, as described in \cite{Rubino2008}, we still lack
an accurate description of the photoionization cross-sections, energies or
transition rates for the Helium atom, which might lead to small changes in these
results.
One may also wonder whether the small differences found for the correction
caused by Helium feedback processes (see Sect.~\ref{sec:feedbackHe}) could
matter.

\begin{figure}
\centering \includegraphics[width=0.9\columnwidth]{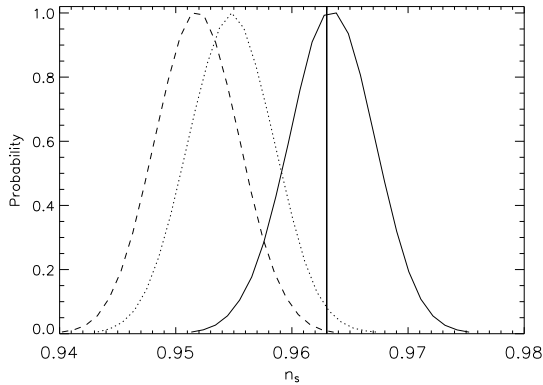}
\caption{Bias on the $\nS$ parameter for different modellings of the Helium
  recombination history. The black solid line corresponds to the posterior
  distribution recovered using the {\sc Rico} code and the additional
  corrections used in Sec.~\ref{sec:additional}. The dotted line corresponds to
  the posterior distribution recovered using the current {\sc Recfast} code
  (v1.4.2), while the dashed line uses the same version of the code but without
  any correction to the helium recombination history. }
\label{fig:ns_oldrecfast}
\end{figure}

By far, the most relevant correction to the helium recombination history is
caused by the absorption of \ion{He}{i} photons by neutral hydrogen, although
the inclusion of the $2^3\rm P_1$-$1^1\rm S_0$ intercombination line also gives
some contribution. These two effects are already included in the codes, both in
{\sc Rico} (FCRW09) and in {\sc Recfast} v1.4.2 \citep{Wong2008}. For
completeness, we have also quantified for this paper the impact that these two
corrections to the Helium recombination would have on the recovered cosmological
parameters.
Our computations show that, if these corrections are not included (which in
practise corresponds to setting {\verb!RECFAST_Heswitch! = 0} in {\sc Recfast}
v1.4.2, or equivalently, using {\sc Recfast} v1.3), then the resulting biases on
the parameters are found to be -3.2, -2.0, -1.2 and -0.7 sigmas for $\nS$,
$\omegab$, $\logAs$ and $H_0$, respectively. For illustration, we show in
Fig.~\ref{fig:ns_oldrecfast} a comparison of the posterior distribution obtained
for the $\nS$ parameter when using the {\sc Recfast} code with
({\verb!RECFAST_Heswitch! = 6}) and without ({\verb!RECFAST_Heswitch! = 0}) all
the corrections to the helium recombination.

This simple computation shows that additional corrections to the helium
recombination history at the $\sim 0.1\%$ level will not matter much to the
analysis of future {\sc Planck} data. Therefore, the physics of helium
recombination already seems to be captured at a sufficient level of precision,
when including the acceleration caused by the hydrogen continuum opacity and the
$2^3\rm P_1$-$1^1\rm S_0$ intercombination line, which together lead to a $\sim
-3\%$ correction to $X_{\rm e}(z)$ at $z\sim 1800$ \citep{rico}.

\subsection{Recombination uncertainties and other extended cosmological models}
\label{sec:nonstandard}

In addition to those extended models described in
Sect.~\ref{sec:extendedmodels}, there is a number of possible non-standard
models for which the inclusion of refined recombination physics might be of
importance.
For example, when using current CMB data to constrain the presence of
hypothetical sources of Ly$\alpha$ resonance radiation or ionizing photons at
high redshifts \citep[e.g.][]{Peebles2000, Bean2003, Bean2007}; or to probe dark
matter models with large annihilation cross-section
\citep[e.g][]{Padmanabhan2005, Galli2009, Huetsi2009}; or energy release by
long-lived unstable particles \citep{Chen2004, Zhang2007};
or when exploring the variation of fundamental constants with time (see
e.g. \cite{Galli2009b} for Newton's gravitational constant, or
\cite{2008PhRvD..78h3527L} for the fine structure constant and the Higgs vacuum
expectation value), it is obvious that neglecting physically well understood
additions to the recombination model, as described in Sect.~\ref{sec:physics},
could lead to {\it spurious detections} or {\it confusion}, in particular, when
the possible effects are already known to be rather small \citep[e.g. see][in
  the case of dark matter annihilations]{Galli2009}.

More generally speaking, given that the largest recombination uncertainties are
obtained for $\nS$, $\omegab$ and $\logAs$, one can say that any additional
parameter showing a strong correlation with those three might be biased if an
incomplete description of the recombination physics is used.
In this sense, neglecting the refinements to the recombination model could be as
important as not taking into account, for instance, uncertainties in the beam
shapes, which also have been shown to compromise our ability to measure $\nS$
\citep{Colombo2009};
or the combined effect of beam and calibration uncertainties, which introduce
significant biases to $\nS$ and $\omegab$ \citep{Bridle2002}, although other
parameters (like $\omegak$) are essentially not affected because these are
basically constrained by the position of the peaks, and not by their amplitudes.

\subsection{Recombination modelling and other cosmological probes}

The combination of CMB data with other datasets usually helps to improve the
parameter constraints, in some cases by breaking internal degeneracies which are
inherent to the CMB data alone. One of the commonly used external datasets is
the Baryon Acoustic Oscillations (BAOs). \cite{deBernardis2009} have recently
shown that a possible delay of recombination \citep{Peebles2000} by extra
sources of ionizing or exciting photons leads to biases on the constraints from
BAOs, because they largely rely on the determination of the size of the acoustic
horizon at recombination. Their conclusions can be directly translated here,
stressing that a fully consistent combination of the constraints from CMB and
BAOs should be done by using the same recombination history in both cases.

In addition, we would like to point out that in principle one could reduce the
uncertainty in our knowledge of the recombination epoch and its possible {\it
  non-standard} extensions \citep[e.g. due to annihilating dark
  matter][]{Padmanabhan2005} in two ways.
On one hand, one could search for the imprint of the cosmological hydrogen
recombination lines on the CMB angular power spectrum
\citep{RHS2005,Carlos2007}, which arises due to the resonant scattering of CMB
photons by hydrogen atoms at each epoch \citep{Basu2004}.
On the other hand, one could also try to {\it directly observe} the photons that
are emitted during the recombination epoch.
Today these photons should still be visible as small distortion of the CMB
energy spectrum in the mm, cm and dm spectral bands
\citep[e.g. see][]{Dubrovich1975, Dubrovich1997, Rubino2006, Chluba2006a,
  Chluba2007, Rubino2008, Chluba2009d}.
These observations not only would open another possibility to determine some of
the key cosmological parameters, such as the {\it primordial helium abundance},
the {\it number density of baryons} and the CMB {\it monopole temperature} at
recombination \citep[e.g.][]{Chluba2008}, but they would also allow us to {\it
  directly} check our understand of the recombination process and possible
non-standard aspects \citep[e.g. see][for an overview]{Sunyaev2009}, for
example, in connection with early energy release \citep{Chluba2008b}, or dark
matter annihilations \citep{Chlubaprep2009}.

\section{Conclusions}
\label{sec:conclusions}

In this paper, we have performed a MCMC analysis of the expected biases on the
cosmological constraints to be derived from the upcoming {\sc Planck} data, in
the light of recent developments in the description of the standard cosmological
recombination process. Our main conclusions are:
\begin{itemize}
\item An incomplete description of the cosmological recombination process leads
  to significant biases (of several sigmas) in some of the basic parameters to
  be constrained by {\sc Planck} satellite (see Table~\ref{tab:planck2}), and in
  general, by any future CMB experiment. However, these corrections have a minor
  impact for present-day CMB experiments; for instance, using WMAP5 data plus
  other cosmological datasets, we find a $\sim -0.25$ and $\sim -0.22$ sigma
  bias on $\nS$ and $\logAs$, respectively, while the rest of the parameters
  remain unchanged.

\item Today, it seems that our understanding of cosmological recombination has
  reached the sub-percent level in $X_{\rm e}$ at redshifts $500 \la z \la
  1600$. However, it will be important to cross-validate all of the considered
  corrections in a detailed code comparison, which currently is under discussion
  among the different groups.
 
\item Given the range of variation of the relevant cosmological parameters, it
  is possible to incorporate all the new recombination corrections by using
  (cosmology independent) fudge functions.
Here we described one possibility which uses a simple correction factor to the
results obtained with {\sc Recfast} (see Sect.~\ref{sec:corr_fac}). We provide
the function $f(z)$ on the {\sc
  Rico}-webpage \footnote{http://cosmos.astro.uiuc.edu/rico}.

\item The physics of helium recombination already seems to be captured at a
  sufficient level of precision, when including the acceleration caused by the
  hydrogen continuum opacity and the $2^3\rm P_1$-$1^1\rm S_0$ intercombination
  line. The biases caused by neglecting only these corrections are -0.8 and -0.4
  sigmas, for $\nS$ and $\omegab$, respectively.

\item When allowing for more non-standard additions to the recombination model
  (e.g. related to annihilating dark matter), the biases introduced by an
  inaccurate recombination model could lead to spurious detections or additional
  confusion (see Sect.~\ref{sec:nonstandard}).

\end{itemize}

\section*{Acknowledgements}
JAR-M and JC are grateful to R.~A.~Sunyaev for useful comments and
discussion. Furthermore, they would like to thank M.~Bucher, J.~Fung, S.~Galli,
D.~Grin, Y.~Haimoud, C.~Hirata, U.~Jentschura, S.~Karshenboim, E.~Kholupenko,
L.~Labzowsky, D.~Scott, and D.~Solovyev for stimulating and very friendly
discussion during the recombination workshop held in July 2009 in Orsay/Paris.
The authors are also very grateful to E.~Switzer for comments and useful
discussions.

This work has been partially funded by project AYA2007-68058-C03-01 of the
Spanish Ministry of Science and Innovation (MICINN). JAR-M is a Ram\'on y Cajal
fellow of the MICINN.
WAF was supported through a fellowship from the Computational Science and
Engineering program at the University of Illinois. Some of the calculations in
this work used the Linux cluster in the University of Illinois Department of
Physics.


\label{lastpage}

\end{document}